\documentclass[journal]{IEEEtran}

\usepackage[margin=1in]{geometry}

\usepackage{amssymb,amsmath,amsthm,amsfonts}
\usepackage{mathtools}
\usepackage{enumitem}
\usepackage[numbers,comma,sort&compress]{natbib}
\usepackage{authblk}
\usepackage{graphicx}
\usepackage[font=small]{caption}

\usepackage[labelformat=simple]{subcaption}
\usepackage{float}
\usepackage[ruled,vlined,linesnumbered]{algorithm2e}
\usepackage{physics}
\usepackage{footnote}
\usepackage{xcolor}
\usepackage{bbding}

\usepackage{tikz}
\usetikzlibrary{calc,patterns,angles,quotes}
\usepackage{siunitx}
\usetikzlibrary{quantikz}

\usepackage{hyperref}

\newtheorem{theorem}{Theorem}
\newtheorem{definition}{Definition}
\newtheorem{lemma}{Lemma}

\newtheorem{corollary}{Corollary}

\newcommand{\eq}[1]{Eq.~(\ref{eq:#1})}

\newcommand{\thm}[1]{\hyperref[thm:#1]{Theorem~\ref*{thm:#1}}}
\newcommand{\cor}[1]{\hyperref[cor:#1]{Corollary~\ref*{cor:#1}}}
\newcommand{\defn}[1]{\hyperref[defn:#1]{Definition~\ref*{defn:#1}}}
\newcommand{\lem}[1]{\hyperref[lem:#1]{Lemma~\ref*{lem:#1}}}
\newcommand{\prop}[1]{\hyperref[prop:#1]{Proposition~\ref*{prop:#1}}}
\newcommand{\fig}[1]{\hyperref[fig:#1]{Figure~\ref*{fig:#1}}}
\newcommand{\tab}[1]{\hyperref[tab:#1]{Table~\ref*{tab:#1}}}
\newcommand{\algo}[1]{\hyperref[algo:#1]{Algorithm~\ref*{algo:#1}}}
\renewcommand{\sec}[1]{\hyperref[sec:#1]{Section~\ref*{sec:#1}}}
\newcommand{\append}[1]{\hyperref[append:#1]{Appendix~\ref*{append:#1}}}
\newcommand{\fac}[1]{\hyperref[fac:#1]{Fact~\ref*{fac:#1}}}
\newcommand{\lin}[1]{\hyperref[lin:#1]{Line~\ref*{lin:#1}}}

\def\>{\rangle}
\def\<{\langle}

\renewcommand{\ket}[1]{|#1\rangle}

\renewcommand{\d}{\mathrm{d}}

\def\:{\hbox{\bf:}}

\def \eps {\epsilon}

\makeatletter
\newcommand{\doubletilde}[1]{{%
  \mathpalette\double@tilde{#1}%
}}
\newcommand{\double@tilde}[2]{%
  \sbox\z@{$\m@th#1\tilde{#2}$}%
  \ht\z@=.9\ht\z@
  \tilde{\box\z@}%
}
\makeatother

\let\oldnl\nl
\newcommand{\nonl}{\renewcommand{\nl}{\let\nl\oldnl}}

\SetKwComment{Comment}{/* }{ */}

\SetKwInOut{Input}{Input}
\SetKwInOut{Registers}{Registers}
\SetKwInOut{Parameters}{Parameters}
\SetKwInOut{Notations}{Notations}
\SetKwInOut{Output}{Output}
\SetKw{Apply}{Apply}
\SetKw{Construct}{Construct}
\SetKw{Estimate}{Estimate}
\SetKw{Return}{Return}

\begin{document}

\title{Improved Quantum Algorithms for R{\'e}nyi Entropies Estimation}

\author{Zongxing Xiong, Chengshen Gao, Lvzhou Li
\thanks{}
}

\maketitle

\begin{abstract}
Entropy estimation of discrete probability distributions is a fundamental problem in statistical inference and quantum property estimation. In this paper, we study the problem of estimating R{\'e}nyi entropy $H_\alpha(\mathbf{p})$ for any distribution $\mathbf{p} = (p_i)_{i=1}^n$ given its purified quantum query-access. For non-integer $\alpha > 1$,  our algorithm estimates $H_\alpha(\mathbf{p})$ to within additive error $\epsilon$ using only $\widetilde{\mathcal{O}}\big(n^{\max\{1/2,1-1/\alpha\}}/\epsilon\big)$ queries, which improves the best known upper bound for dependence on both $n$ and $1/\epsilon$. For $0<\alpha<1$, our algorithm has query complexity $\widetilde{\mathcal{O}}\big(n^{1/(2\alpha)}/\epsilon^{\max\{1,1/(2\alpha)\}}\big)$, which improves the dependence on $1/\epsilon$ of the current best upper bound without incresing its $n$ dependence. Our approach adapts the multi-level estimation framework to relative approximation of the power sums and leverage binary search to remove the requirement of prior knowledge of them. We also strengthen the precision-dependent lower bound to $\Omega\big(1/\epsilon^{\max\{1,1/(2\alpha)\}}\big)$ for $0<\alpha<1$, which identifies the optimal precision dependence of our algorithm.
\end{abstract}



\section{Introduction}

Estimating numerical properties of an unknown probability distribution is a fundamental problem in statistics and theoretical computer science.
Rather than reconstructing the entire distribution, the goal is to estimate a specified functional to a prescribed accuracy using as few samples or oracle queries as possible.

An important family of such functionals is entropy, which quantifies uncertainty and is a basic tool in information theory, statistics, and theoretical computer science. Shannon entropy characterizes the fundamental limit of lossless data compression~\cite{Shannon,coverthomas}, while R{\'e}nyi entropies provide a family of measures that distinguish different features of a distribution~\cite{Renyi}. R{\'e}nyi entropies arise in the analysis of guessing moments~\cite{arikan}, and their smooth variants have operational applications to randomness extraction and privacy amplification~\cite{rennerwolf}. Estimating these quantities therefore provides information that is not captured by a single entropy value.

When the distribution is unknown, computing its entropy is a statistical problem. A direct approach first estimates each probability and then substitutes these estimates into the entropy formula. For a large alphabet, however, many outcomes may be observed rarely or not at all, and the resulting bias can be substantial~\cite{paninski}. Estimating a scalar functional can require fewer samples than learning the entire distribution, as demonstrated by sublinear entropy estimators and polynomial approximation methods~\cite{valiantvaliant,jiao,wuyang}. The R{\'e}nyi family is particularly instructive: its classical sample complexity changes markedly with $\alpha$, and integer orders above one admit collision-based estimators that are unavailable for general non-integer orders~\cite{acharya,obremski}. These differences make non-integer R{\'e}nyi entropy a natural problem for understanding the role of quantum access in statistical estimation.

Quantum algorithms offer a way to estimate such properties using coherent access to the underlying distribution.
We consider purified quantum query-access to $\mathbf{p}$, where a unitary $U_p$ prepares
\begin{equation*}
U_p\ket{0}_G\ket{0}_B
=\sum_{i=1}^n\sqrt{p_i}\ket{\phi_i}_G\ket{i}_B,
\end{equation*}
and its inverse and controlled versions are also available. The normalized garbage states $\ket{\phi_i}$ need not be known. Measuring the label register produces an ordinary sample, while coherent access permits interference between successive oracle calls. This model encompasses reversible sampling procedures and supports distributional property estimation through projected unitary encodings~\cite{gl}. Our resource measure is the number of oracle queries; the cost of constructing a preparation oracle is not included. The objective is to output an additive $\epsilon$-approximation to $H_\alpha(\mathbf{p})$ with prescribed success probability, without learning all the probabilities.

Quantum algorithms for distributional property testing and entropy estimation \cite{bhh,gl,montanaro,luo,CKO25,liwu,wzl,multilevel} have been developed using amplitude estimation \cite{AE} and quantum singular value transformation (QSVT) \cite{qsvtstoc}.
In particular, Wang, Zhang, and Li \cite{wzl} introduced a framework combining QSVT, quantum annealing, and variable-time amplitude estimation. Their algorithms estimate R{\'e}nyi entropy using
\begin{equation}
    \widetilde{\mathcal{O}}\!\left(
        \frac{n^{1-1/(2\alpha)}}{\epsilon}
        +\frac{\sqrt n}{\epsilon^{1+1/(2\alpha)}}
    \right)
\end{equation}
queries for $\alpha>1$, and
\begin{equation}
    \widetilde{\mathcal{O}}\!\left(
        \frac{n^{1/(2\alpha)}}{\epsilon^{1+1/(2\alpha)}}
    \right)
\end{equation}
queries for $0<\alpha<1$. Their use of different transformation polynomials on different intervals addresses the varying approximation costs associated with probabilities of different magnitudes. More recently, Chen et al. \cite{multilevel} introduced a multilevel framework for estimating functionals of discrete distributions and obtained improved algorithms for Tsallis entropy. For R{\'e}nyi entropy with $0<\alpha<1$, their results, combined with those of Wang et al.~\cite{wzl},
give the query upper bound
\begin{equation}
    \widetilde{\mathcal{O}}\!\left(
        \min\left\{
            \frac{n^{1/(2\alpha)}}{\epsilon^{1+1/(2\alpha)}},
            \frac{n^{1/\alpha-1/2}}{\epsilon^{1/\alpha}}
        \right\}
    \right).
\end{equation}
These results leave open whether the quantum query complexity of R{\'e}nyi entropy estimation can be further improved,
particularly in its dependence on both the distribution size and the precision $\epsilon$.

A central issue is the accuracy with which the power sum must be estimated. Although R{\'e}nyi and Tsallis entropies are both determined by $P_\alpha(\mathbf{p})$, their estimation requirements differ. Additive estimation of Tsallis entropy requires additive accuracy in the power sum. By contrast, for fixed $\alpha\neq1$, additive estimation of R{\'e}nyi entropy requires relative accuracy in the power sum. Indeed, if
$|\widetilde P_\alpha-P_\alpha|\leq rP_\alpha$ with $r\leq1/2$, then
\begin{equation}
    \left|
        \frac{\log\widetilde P_\alpha-\log P_\alpha}{1-\alpha}
    \right|
    \leq \frac{2r}{|1-\alpha|}.
\end{equation}
Consequently, the magnitude of $P_\alpha(\mathbf{p})$ is directly relevant to the estimation cost. This suggests the following question:
\begin{quote}
Can the multilevel framework exploit the magnitude of the power sum
to improve the query complexity of R{\'e}nyi entropy estimation?
\end{quote}

\subsection{Main results}

We give R{\'e}nyi entropy estimators with improved query complexity for any fixed non-integer $\alpha>1$ and for $0<\alpha<1$, together with a strengthened lower bound in the latter regime. Throughout the following discussions, $\widetilde{\mathcal{O}}$ hides poly-logarithmic factors in $n$, $1/\epsilon$, and $1/\delta$, where $\delta$ is the maximum failure probability. 

\textit{1) Noninteger orders above one:} In \thm{non-integer-large}, we obtain the query upper bound
\begin{equation*}
\widetilde{\mathcal{O}}\!\left(
\frac{n^{\max\{\frac12,1-\frac1\alpha\}}}{\epsilon}
\right)
\end{equation*}
for estimating R{\'e}nyi entropy $H_\alpha(\mathbf{p})$ for non-integer $\alpha > 1$.
Thus, the complexity is $\widetilde{\mathcal{O}}(\sqrt n/\epsilon)$ for $1<\alpha<2$ and $\widetilde{\mathcal{O}}(n^{1-1/\alpha}/\epsilon)$ for non-integer $\alpha>2$. The corresponding best known upper bound is
\begin{equation*}
\widetilde{\mathcal{O}}\!\left(
\frac{n^{1-1/(2\alpha)}}{\epsilon}
+\frac{\sqrt n}{\epsilon^{1+1/(2\alpha)}}
\right)
\end{equation*}
given in ~\cite{wzl}. Our result lowers the exponent of $n$ in the first term and achieves linear dependence on $1/\epsilon$ throughout the $\alpha>1$ regime. For $1<\alpha<2$ and constant accuracy, the dimensional exponent matches the lower bound $\Omega(n^{1/2-o(1)})$ from quantum sample-to-query lifting~\cite{lifting}, up to subpolynomial factors.

\textit{2) Orders below one:} In \thm{non-integer-small}, we obtain query upper bound
\begin{equation*}
\widetilde{\mathcal{O}}\!\left(
\frac{n^{1/(2\alpha)}}
{\epsilon^{\max\{1,1/(2\alpha)\}}}
\right)
\end{equation*}
for estimating R{\'e}nyi entropy $H_\alpha(\mathbf{p})$ for $0 < \alpha < 1$. Compared with the the current best upper bound $$\widetilde{\mathcal{O}}\!\left(\min\left\{\frac{n^{1/(2\alpha)}}{\epsilon^{1+1/(2\alpha)}},\frac{n^{1/\alpha-1/2}}{\epsilon^{1/\alpha}}\right\}\right)
$$ 
given in \cite{multilevel}, ours improves the dependence on $1/\epsilon$ without increasing the exponent of $n$. There is a transition at $\alpha=1/2$: the complexity is $\widetilde{\mathcal{O}}((n/\epsilon)^{1/(2\alpha)})$ for $0<\alpha\leq1/2$, and $\widetilde{\mathcal{O}}(n^{1/(2\alpha)}/\epsilon)$ for $1/2\leq\alpha<1$.

\textit{3) Precision-dependent lower bound:} For fixed $0<\alpha<1$, \thm{non-integer-small-lower} gives a stronger query lower bound
\begin{equation*}
\Omega\!\left(
n^{\frac{1}{2\alpha}-o(1)}
+\frac{n^{\frac{1}{2\alpha}-\frac12}}
{\epsilon^{\max\{1,1/(2\alpha)\}}}
\right)
\end{equation*}
than the current best of \cite{lifting}. This follows by perturbing the nonzero tail mass of a distribution with one heavy outcome and a uniform light tail. Two such distributions have an entropy gap exceeding $2\epsilon$, while suitable purification oracles are close in operator norm. A hybrid argument then gives $\Omega(n^{1/(2\alpha)-1/2}/\epsilon)$ queries. Combining this with the original bound of~\cite{lifting} strengthens the precision-dependent term for $1/2<\alpha<1$ and identifies the optimal precision dependence of our algorithm, up to logarithmic factors.

\subsection{Related works}
\textit{Classical entropy and functional estimation:} Paninski~\cite{paninski} analyzed the bias and consistency of entropy estimators, including the difficulties caused by undersampling. Valiant and Valiant~\cite{valiantvaliant} established that entropy can be estimated with a sublinear number of samples in the alphabet size. Jiao, Venkat, Han, and Weissman~\cite{jiao} and Wu and Yang~\cite{wuyang} developed minimax estimators based on polynomial approximation, clarifying the distinct roles of bias reduction and statistical fluctuations. Related approximation methods also yield optimal support-size estimators~\cite{supportsize}. These works show why direct estimation of a functional can be preferable to reconstructing an unknown distribution.

For R{\'e}nyi entropy, Acharya, Orlitsky, Suresh, and Tyagi~\cite{acharya} identified different sample-complexity regimes for orders below one, non-integer orders above one, and integer orders above one. Obremski and Skorski~\cite{obremski} developed entropy-dependent bounds for integer orders and lower bounds with explicit alphabet and accuracy dependence. At constant accuracy, the characteristic classical dimensional exponents are $1/\alpha$ for $0<\alpha<1$ and one for non-integer $\alpha>1$, up to subpolynomial factors. The integer case is special because $P_\alpha$ is a collision probability. Our work addresses the non-integer regimes using coherent query access, a stronger resource than independent samples.

\textit{Quantum distributional property estimation:} Bravyi, Harrow, and Hassidim~\cite{bhh} gave early quantum algorithms for testing and comparing distributions. Montanaro and de Wolf~\cite{testingsurvey} surveyed the broader development of quantum property testing, and Belovs~\cite{belovs} characterized quantum distribution discrimination in several access models. Li and Wu~\cite{liwu} studied the quantum query complexity of entropy estimation and established speedups for Shannon and R{\'e}nyi entropies. Gily{\'e}n and Li~\cite{gl} developed methods for distributional property testing under purified access. These access-model distinctions matter: bounds based on independent samples, coherent evaluation of a classical sampler, and general purification oracles must be interpreted with their corresponding assumptions.

The algorithmic primitives underlying this line of work include amplitude amplification and estimation~\cite{AE}, quantum Monte Carlo estimation~\cite{montanaro}, and quantum mean estimation~\cite{hamoudi}. Quantum signal processing~\cite{lowchuang} and quantum singular value transformation~\cite{qsvtstoc,qsvtfull} make polynomial transformations available within quantum circuits. Variable-time amplitude amplification~\cite{ambainis} provides another way to account for branches with different costs. Wang, Zhang, and Li~\cite{wzl} combined singular value transformation, quantum annealing, and variable-time amplitude estimation to obtain a framework for R{\'e}nyi entropy estimation. Their bounds provide the direct quantitative comparison for the results above.

Chen, Gao, Li, Wang, and Wang~\cite{multilevel} introduced the multi-level framework used here and applied it to functionals including Tsallis entropy. It combines geometric probability intervals with non-destructive singular value discrimination and local polynomial approximation. Although Tsallis and R{\'e}nyi entropies both depend on power sums, an additive approximation to a power sum need not have the relative accuracy required after taking its logarithm. Our contribution is to adapt this framework to that relative-accuracy requirement, exploiting the power-sum scale for $\alpha>1$ and changing the initial state for $0<\alpha<1$.

\textit{Quantum lower bounds:} Wang and Zhang~\cite{wangzhanglifting} established quantum sample-to-query lifting, and Tang, Wright, and Zhandry~\cite{twz} extended sample-based simulation to state-preparation oracles. Chen, Wang, and Zhang~\cite{lifting} applied these developments to obtain lower bounds for distributional properties, including R{\'e}nyi entropy. We use their dimensional and small-order precision bounds as benchmarks and supplement them with a direct two-oracle argument. This separates the improvement proved here from the lower bounds inherited from previous work.

\subsection{Organization}
In \sec{prelim}, we introduce the access model, projected unitary encodings, polynomial approximations, and estimation primitives. In \sec{non-integer-large}, we present the algorithm for non-integer $\alpha>1$, first with a known power-sum bound and then with geometric search. In \sec{non-integer-small}, we treat $0<\alpha<1$ and prove the strengthened lower bound.


\section{Preliminaries}\label{sec:prelim}

\subsection{Notation}

For an integer $n\ge2$, write $[n]=\{1,\ldots,n\}$. Let $\mathbf{p}=(p_1,\ldots,p_n)$ be a probability distribution, allowing zero entries. Among statistical properties, the most basic one is the Shannon entropy~\cite{Shannon}, which is defined as
\begin{align}
    H(\mathbf{p}):=-\sum_{i=1}^{n}p_{i}\log p_{i}.
\end{align}A natural generalization of the Shannon entropy is the family of R{\'e}nyi entropies~\cite{Renyi}. Specifically, the $\alpha$-R{\'e}nyi entropy is defined as
\begin{align}
    H_\alpha(\mathbf{p}) := \frac{1}{1-\alpha} \log \sum_{i=1}^n p_i^\alpha.
\end{align}
For our convenience, the power sum in the logarithm is denoted by $P_\alpha(\mathbf{p})$, i.e., $P_\alpha(\mathbf{p}) := \sum_{i=1}^n p_i^\alpha$. When $\alpha\to 1$, $\lim_{\alpha\to 1} H_{\alpha}(\mathbf{p})=H(\mathbf{p})$.
Throughout this paper, logarithms have natural bases unless otherwise stated and all vector norms $\|\cdot \|$ in this paper are $L^2$-norm.

\subsection{Purified quantum query-access}
\begin{definition}[Purified Quantum Query-Access]\label{defn:purified-access}
    A discrete distribution $\mathbf{p}=(p_i)_{i=1}^n$ has purified quantum query-access if we have access to a unitary oracle $U_p$ (and its inverse and  controlled versions) acting as 
    
    \begin{equation*}
        U_p | 0 \rangle_G | 0 \rangle_B = \sum_{i-1}^n \sqrt{p_i} | \phi_i \rangle_G | i \rangle_B,
    \end{equation*}
    where $\{| \phi_i \rangle \colon i \in [n] \}$ are arbitrary unknown garbage states.
\end{definition}

\subsection{Projected unitary encoding}
To manipulate general matrices $A$ by quantum circuits, we need a tool called \textit{projected unitary encoding} introduced by \cite{qsvtstoc}.
A projected unitary encoding of an operator $A$ is a triple $(\widetilde\Pi,\Pi,U)$ with $A=\widetilde\Pi U\Pi$, where $\Pi$ and $\widetilde\Pi$ are orthogonal projections. 

Here we list some useful projected unitary encodings intrduced by previous work and will be used in this article.

\begin{itemize}
\item For $U_p$ in \defn{purified-access}, add a label register $C$, set $U=U_{p} \otimes I_C$, $\Pi=|0\rangle\langle 0|_G \otimes |0\rangle\langle 0|_B \otimes I_C$ and $\widetilde{\Pi}=\sum_{i=1}^{n} I_G \otimes|i\rangle\langle i|_B \otimes| i\rangle\langle i|_C$, then we have
    \begin{align}\label{eq:encoding1}
     \widetilde{\Pi} U \Pi=\sum_{i=1}^{n} \sqrt{p_{i}}\left|\phi_{i}\right\rangle\langle 0|\otimes| i\rangle\langle 0|\otimes| i\rangle\langle i|.
    \end{align}
\item For $U_p$ in \defn{purified-access}, add a label register $C$ and a qubit $D$, let $R_D |0\rangle =\frac12 |0\rangle + \frac{\sqrt{3}}{2} |1\rangle$, then set $U = U_p^\dagger \otimes I_C \otimes R_D^\dagger$, $\Pi = \sum_i I_G \otimes |ii\rangle\langle ii|_{BC} \otimes |0\rangle\langle 0|_D$ and   $\widetilde{\Pi}=|00\rangle\langle 00|\otimes I_C \otimes |0\rangle\langle 0|_D$, then we have
    \begin{equation}\label{eq:encoding2}
     \widetilde{\Pi} U \Pi = \sum_i\frac{\sqrt{p_i}}{2} |l_i\rangle\langle r_i|, 
     \begin{cases}
     |l_i\rangle = |0,0,i,0\rangle, \\
     |r_i\rangle = |\phi_i,i,i,0\rangle.
     \end{cases}
    \end{equation}
\item Hermitian dilation of the projected unitary encoding $A=\widetilde\Pi U\Pi$: ~Let $U_H = |0\rangle\langle 1|_E \otimes U^{\dagger} + |1\rangle\langle 0|_E \otimes U$ acting on the original system and an additional auxiliary qubit $E$. With respect to the projector $\Pi_H = |0\rangle\langle 0|_E \otimes \Pi + |1\rangle\langle 1|_E \otimes \widetilde{\Pi}$, we get the projected unitary encoding
    \begin{align}\label{eq:encoding3}
       \Pi_H U_H \Pi_H = H = \begin{bmatrix}
            0 & A^{\dagger}\\
            A & 0
        \end{bmatrix}.
    \end{align}
    The spectral decomposition of $H$ takes the form 
    \begin{equation}\label{eq:hermitian-dilation}
    H = \sum_{i} \sigma_i \big( \ketbra{\phi_{i,+}}{\phi_{i,+}} - \ketbra{\phi_{i,-}}{\phi_{i,-}} \big),
    \end{equation} 
    where the eigenvectors are given by $\ket{\phi_{i,\pm}} = \frac{1}{\sqrt{2}}(\ket{0}\ket{\psi_i} \pm \ket{1}\ket{\tilde{\psi}_i})$.
\end{itemize}

The first projective unitary encoding was proposed by \cite{gl} and will be used in \sec{integer} for estimating R{\'e}nyi entropy with integer $\alpha$. The second one and the third one, proposed by \cite{multilevel}, will be used in \sec{non-integer-large} and \sec{non-integer-small} for estimating R{\'e}nyi entropy with non-integer $\alpha$.

\subsection{Quantum singular value transformation}
In \cite{qsvtstoc}, a general quantum algorithm framework called \textit{quantum singular value transformation (QSVT)} is proposed, which is useful in many computational tasks including property estimation. Before introducing this framework, we first give the definition of singular value transformation.
\begin{definition}[Singular value transformation {\cite[Definition 9]{qsvtstoc}}]
   Let $f\colon \mathbb{R} \rightarrow \mathbb{C}$ be an even or odd function. Suppose that $A \in \mathbb{C}^{\tilde{d} \times d}$ has the following singular value decomposition
  \begin{align}
    A=\sum_{i=1}^{d_{\min }} \sigma_{i}|\tilde{\psi}_{i}\rangle\left\langle\psi_{i}\right|,
  \end{align}
  where $d_{\min }:=\min (d, \tilde{d})$. For the function $f$ we define the singular value transform of $A$ as
  \begin{align}f^{(S V)}(A):= \begin{cases}\sum_{i=1}^{d_{\min }} f\left(\sigma_{i}\right)|\tilde{\psi}_{i}\rangle\left\langle\psi_{i}\right| & \text { if } f \text { is odd, and } \\ \sum_{i=1}^{d} f\left(\sigma_{i}\right)\left|\psi_{i}\right\rangle\left\langle\psi_{i}\right| & \text { if } f \text { is even}, \end{cases}\end{align}
  where for $i \in[d] \backslash\left[d_{\min }\right]$ we define $\sigma_{i}:=0$.
\end{definition}

Given a matrix $A$ block-encoded in a unitary, polynomial singular value transformation of $A$ can be efficiently implemented as follows:
\begin{theorem}[{\cite[Corollary 11]{qsvtstoc}}] \label{thm:qsvt}  Let $\mathcal{H}_{U}$ be a finite-dimensional Hilbert space and let $U, \Pi, \widetilde{\Pi} \in \operatorname{End}\left(\mathcal{H}_{U}\right)$ be linear operators on $\mathcal{H}_{U}$ such that $U$ is a unitary, and $\widetilde{\Pi}, \Pi$ are orthogonal projectors. Suppose that $P=\sum_{k=0}^{n} a_{k} x^{k} \in \mathbb{R}[x]$ is a degree-n polynomial such that
  \begin{align}
  \label{eq:qsvt_condition}
    &a_{k} \neq 0 \text{ only if }k \equiv n \bmod 2\text{, and }\nonumber \\&\text{for all }x \in[-1,1]:|P(x)| \leq 1. 
  \end{align}
  Then there exists a vector $\Phi  = (\phi_1, \phi_2, \ldots, \phi_n)  \in \mathbb{R}^{n}$, such that
  \begin{align}
    &P^{(SV)}\big(\widetilde{\Pi} U \Pi\big) \nonumber \\
    = & \begin{cases}
    \big(\langle+| \otimes \widetilde{\Pi}\big) U^{(SV)}_P \big(|+\rangle \otimes \Pi\big) & \text { if } n \text { is odd,} \\
    \big(\langle+| \otimes \Pi\big) U^{(SV)}_P \big(|+\rangle \otimes \Pi\big) & \text { if } n \text { is even, }\end{cases}
  \end{align}
  where $U^{(SV)}_P := |0\rangle\langle 0|\otimes U_{\Phi}+| 1\rangle\langle 1| \otimes U_{-\Phi}$ with
  \begin{equation}
  U_{\Phi}:=\begin{cases}
    e^{i \phi_{1}(2 \widetilde{\Pi}-I)} U \prod_{j=1}^{(n-1) / 2}(e^{i \phi_{2 j}(2 \Pi-I)} U^{\dagger}\\ \cdot e^{i \phi_{2 j+1}(2 \widetilde{\Pi}-I)} U) \qquad \text {if } n \text { is odd,} \\
    \prod_{j=1}^{n / 2}(e^{i \phi_{2 j-1}(2 \Pi-I)} U^{\dagger} \\\cdot e^{i \phi_{2 j}(2 \widetilde{\Pi}-I)} U) \qquad \  \text { if } n \text { is even. }
    \end{cases}
    \end{equation} 
\end{theorem}
Note that $P^{(S V)}\big(\widetilde{\Pi} U \Pi\big)$ acts on the same space as $\widetilde{\Pi} U \Pi$, while $U_P^{(SV)}$ acts on an enlarged space with one ancillary qubit added. This theorem tells us that for a polynomial $P$ of degree $d$ which satisfies \eq{qsvt_condition} in \thm{qsvt}, we can implement $P^{(SV)}(\widetilde{\Pi}U{\Pi})$ with $\Theta(d)$ uses of $U, U^{\dagger}$ and controlled reflections $I-2\Pi, I-2\widetilde{\Pi}$.

\subsection{Polynomial approximations}

\begin{lemma} [Polynomial approximations of the sign function, {\cite[Lemma 25]{qsvtfull}}] 
\label{lem:sign}
For $\delta > 0$ and $\epsilon \in (0, 1/2)$, there exists an efficiently computable odd polynomial $Q \in \mathbb{R}[x]$ of degree $\mathcal{O}\!\left(\frac{\log(1/\epsilon)}{\delta}\right)$ such that
\begin{itemize}
    \item for all $x \in [-2, 2]$: $|Q(x)| \leq 1$, 
    \item for all $x \in [-2, 2] \backslash (-\delta, \delta)$: $|Q(x) - \mathrm{sign}(x)| \leq \epsilon$.
\end{itemize}
\end{lemma}

\begin{lemma}[Polynomial Approximations of Positive Power Functions, {\cite[Lemma 6]{wzl}}]
\label{lem:positive-powers}
For any $c > 0$, $\beta \in (0,1)$, $\nu \in (0,\beta)$, and $\epsilon \in \left(0,\tfrac{1}{2}\right)$, let
$ g(x) = 2^{-c-1}\beta^{-c} x^{c}$.
There exists an efficiently computable even polynomial $S_{c,\nu,\beta,\epsilon} \in \mathbb{R}[x]$ of degree $
\mathcal{O}\!\left(\frac{c+1}{\nu}\log(\frac{1}{\beta \nu \epsilon})\right) $,
such that
\begin{itemize}
    \item for all $x \in [-1,1]$: $|S(x)| \leq 1$,
    \item for all $x \in [0,\nu]$: $|S(x)| \leq 2g(x)$, 
    \item for all $x \in [\nu,\beta]$: $|S(x) - g(x)| \leq \epsilon$.
\end{itemize}
\end{lemma}

\subsection{Amplitude amplification}
\begin{theorem} [Amplitude estimation {\cite[Theorem 12]{AE}}] \label{AE}
    There is a quantum algorithm that takes as input a unitary $U$, an orthogonal projector $\Pi$, and a positive integer $t$. The algorithm outputs an estimate $\tilde{a}$ of $a = \|\Pi U\ket{0}\|^2$, such that \begin{align*}
        |\tilde{a}-a|\le \frac{2\pi\sqrt{a(1-a)}}{t} + \frac{\pi^2}{t^2}
    \end{align*}
    with probability at least $8/\pi^2$, using $2t+1$ calls to (controlled) $U$ and $U^{\dagger}$, and $t$ calls to the reflection $I-2\Pi$.
\end{theorem}

\begin{corollary} {\cite[Corollary 2.2]{multilevel}}
    \label{cor:two-stage-ae}
    Let $U$ be a unitary and $\Pi$ be an orthogonal projector. Let $a = \|\Pi U\ket{0}\|^2$. For any precision $\delta \in(0,1]$ and failure probability $\eta \in (0, 1)$, there exists a quantum algorithm that outputs an estimate $\tilde{a}$ satisfying $|\tilde{a}-a| \le \eps$ with probability at least $1-\delta$, using 
    \begin{align*}
        O\left( \left(\frac{\sqrt{a}}{\eps} + \frac{1}{\sqrt{\eps}}\right) \log(\frac{1}{\delta}) \right)
    \end{align*}
    queries to (controlled) $U$, $U^\dagger$, and $I-2\Pi$.
\end{corollary}

\subsection{Multi-level estimation framework}
The multi-level framework of \cite{multilevel} provides a general resolution to the problem of estimating functionals $\sum_{i=1}^n f(p_i)$ of a discrete distribution $\mathbf{p}=(p_i)_{i=1}^n$.  Roughly, the framework partitions the interval $[0 ,1]$ into sections
$$[0, \varphi_m], (\varphi_{m}, \varphi_{m-1}],\ldots, (\varphi_1, \varphi_0 = 1]$$ 
with geometrically decreasing $\varphi_j$. In each section $(\varphi_{j}, \varphi_{j-1}]$, a fine-grained polynomial $P_j$ is used to approximate the target function locally rather than the whole interval $[0, 1]$ (in fact a cutoff $[\delta, 1]$ to exclude the singularity) as what standard QSVT frameworks did. Since 

To be more explcit, the authors of \cite{multilevel} rewrite the functional being estimated as $\sum_{i=1}^n p_ig(p_i)$ and they first prepare an initial state in the workspace $\mathsf{w} = EGBCD$:
\begin{equation}\begin{split}\label{eq:prepare}
U_I |\mathbf{0}\rangle_\mathsf{w} & = \sum_{i=1}^n \sqrt{p_i}  |0\rangle_E |r_i\rangle_{GBCD} \\
& = \sum_{i=1}^n \sqrt{\frac{p_i}{2}} \left(|\phi_{i,+} \rangle + |\phi_{i,-} \rangle \right),
\end{split}\end{equation}
with $|r_i \rangle$ being right singular vectors of the projective unitary enconding defined in \eq{encoding2} and $|\phi_{i,\pm} \rangle$ being the eigenvectors of its Hermitian dilation. The preparation unitary can be implemented as 
\begin{equation}\label{eq:prepare-unitary}
U_I = \left( I_{EG} \otimes U_{\mathrm{copy}} \otimes I_D\right) \left(I_E \otimes U_p \otimes I_{CD} \right)
\end{equation}
where $U_{\mathrm{copy}}$ is the bit-wise CNOT gate satisfying $U_{\mathrm{copy}}: |i\rangle_B |\mathbf{0}\rangle_C \mapsto |i\rangle_B |i\rangle_C$. Then, they coherently isolate those components $i$ with $\sqrt{p_i} \in (\varphi_j, \varphi_{j-1}]$ into a target subspace with following non-destructive singular value discriminator:
\begin{theorem} [non-destructive sigular value discrimination {\cite[Theorem 3.1]{multilevel}}]
\label{thm:nd-svd}
    Let $(\widetilde{\Pi}, \Pi, U)$ be a projected unitary encoding of $A$ with spectral norm $\|A\|\le 1/2$. Let $A = \sum_{i=1}^d \sigma_i |\widetilde{\psi_i}\rangle \langle\psi_i|$ be the singular value decomposition of $A$. Let $\ket{\phi_{i,\pm}}$ be the eigenvector of the Hermitian dilation $H$ of $A$. For any precision $\eps_1, \eps_2 \in (0,1]$, threshold $\gamma\in (0, 1/2)$, and constant gap ratio $\rho>1$, there exists a unitary $D$ that maps \begin{align*}  \ket{++} \ket{0}\ket{\phi_{i,\pm}}\mapsto \ket{++}\ket{\xi_{i,\pm}}\ket{\phi_{i,\pm}}
    \end{align*}
   with precision $\eps_1$ for all $i\in [d]$, where $\ket{\xi_{i,\nu}}$ satisfies 
   \begin{align*}
        \begin{cases}
        \|\ket{\xi_{i,\pm}}\mp \ket{0}\|\le \eps_2, &  \text{ if }\sigma_i\in [\gamma, 1),\\
         \|\ket{\xi_{i,\pm}} - \mathrm{i} \ket{1}\|\le \eps_2 , & \text{ if }\sigma_i \in [0, \frac{\gamma}{\rho}]. \\
        \end{cases} 
    \end{align*}
    Implementing $D$ requires $\mathcal{O}(\log(1/\eps_1)+\log(1/\eps_2)/\gamma)$ queries to controlled $U$ and $U^{\dagger}$.
\end{theorem}
Due to the discontinuity of the threshold function, only a “soft” thresholding with uncertain transition region can be implemented. This is settled in \cite{multilevel} by applying two successive thresholding steps $D_{j-1}$ and $D_j$ corresponding to the lower and upper boundaries. After that, the polynomial singular value transformation 
\begin{equation}\begin{split}
P_{j}^{(\mathrm{SV})}(\Pi_H U_H \Pi_H) = & \sum_{i} P_j\left(\frac{\sqrt{p_i}}{2}\right) \ketbra{\phi_{i,+}}{\phi_{i,+}} \\
& + P_j\left(-\frac{\sqrt{p_i}}{2}\right) \ketbra{\phi_{i,-}}{\phi_{i,-}}
\end{split}\end{equation}
is applied, with a local polynomial $P_j$ satisfying $P_j\left(\frac{\sqrt{p_i}}{2}\right)^2 \approx g(p_i)$ (up to some normalization factor) for $\sqrt{p_i} \in (\varphi_{j+1}, \varphi_{j-1}]$. Finally, perform quantum amplitude estimation to measure the squared norm of the state in the isolated target subspace and sum up the outputs of the repeated procedures for every $j \in [m]$, a well approximation of the funtional $\sum_{i=1}^n f(p_i)$ can be achieved.


\section{NON-INTEGER R{\'E}NYI ENTROPY ESTIMATOR ($\alpha>1$)}\label{sec:non-integer-large}

\begin{theorem}\label{thm:non-integer-large}
  For any fixed $\alpha > 1$, there exists an algorithm $\mathcal{A}$ such that for any $\delta \in (0,1)$, $\epsilon\in(0,1)$ and probability distribution $\mathbf{p}$ on $[n]$, $\mathcal{A}$ can estimate $H_{\alpha}(\mathbf{p})$ to within additive error $\epsilon$
  with success probability at least $1-\delta$ using 
  \begin{equation}\label{eq:non-integer-large-complexity}
  \widetilde{\mathcal{O}}\!\left(\frac{n^{\max\{\frac12,1-\frac{1}{\alpha}\}}}{\epsilon}\right)
  \end{equation}
  queries to $U_{p}$ and $U_{p}^{\dagger}$ in \defn{purified-access}.
\end{theorem}
For $1<\alpha<2$, the query upper bound is $\widetilde{\mathcal{O}}(\sqrt{n}/\epsilon)$; for $\alpha > 2$, it is $\widetilde{\mathcal{O}}(n^{1-1/\alpha}/\epsilon)$. Compared with the best known upper bound $\widetilde{\mathcal{O}}\!\left(\frac{n^{1-1/(2\alpha)}}\epsilon + \frac{\sqrt n}{\epsilon^{1+1/(2\alpha)}}\right)$ of \cite[Theorem 7]{wzl}, our algorithm achieve polynomial improvement on both the dimensional dependence and the precision dependence. For $1<\alpha<2$ and constant accuracy, the dimensional exponent matches the lower bound $\Omega(n^{1/2-o(1)})$ from \cite[Theorem 2.8]{lifting}, up to subpolynomial factors. 

We first construct a weaker algorithm which can estimate $P_{\alpha}(\mathbf{p})$ to within multiplicative accuracy $\eta < 1/4$ given its upper bound, and then remove the requirement of knowing an upper bound.

\begin{algorithm*}[ht]
\caption{Quantum algorithm for estimating $P_{\alpha}(\mathbf{p})$ (non-integer $\alpha > 1$) from its known upper bound}
\label{algo:algorithm3}
\Input{Upper bound $R \leq 1$; Multiplicative accuracy $\eta \in (0, 1/4)$; Maximum failure probability $\delta \in (0, 1/3)$;
    Purified access oracle $U_p$ for the distribution $\mathbf{p} = (p_i)_{i=1}^n$\;
}
\Registers{Auxiliary qubits $\mathsf{a}, \mathsf{b}, \mathsf{c}, \mathsf{d}$ and workspace $\mathsf{w}$ defined in \eq{prepare}\;
}
\Parameters{$\Delta := \eta R$; $m := \lceil\frac{1}{2\alpha}\log_2(\frac{12nR}{\Delta})\rceil$; Intervals $(\varphi_{m+1}, \varphi_m], (\varphi_m, \varphi_{m-1}], \cdots, (\varphi_1, \varphi_0]$ with $\varphi_0 := R^{1/2\alpha}$ and $\varphi_j := \varphi_0 / 2^{j}$ for $j \in [m+1]$; $B_{m+1}:=0$ and $B_j := 2^{2\alpha}\varphi_{j-1}^{2\alpha-2}$, local polynomial $P_j := S_{c,\nu,\beta,\epsilon}$ with $c := \alpha-1, \nu := \frac{\varphi_{j+1}}{2}, \beta := \frac{\varphi_{j-1}}{2}, \epsilon := \frac{\Delta}{36B_j}$ in \lem{positive-powers}, $\overline{v}_j := \min\big\{1, 5n\varphi_{j-1}^2/4, 20R/\varphi_{j+1}^{2\alpha - 2}\big\}$ for each $j \in [m]$\;
}
\Output{
    Estimate $\widetilde{P}_\alpha(\mathbf{p})$ satisfying \eq{algorithm3-output}.
}

\BlankLine
$S \leftarrow 0$\;
Let $(\Pi_H, \Pi_H, U_H)$ be the Hermitian dilation of the projected unitary encoding defined in \eq{encoding2} and $U_I$ be the state preparation unitary defined in \eq{prepare-unitary}\;
Let $D_0 = I$ and for $j\in [m]$, let $D_j$ be the non-destructive singular value discriminator $D$ in \thm{nd-svd} with threshold $\gamma = \varphi_j/2$, gap ratio $\rho = 2$, and precision $\eps_{j,1} = \Delta/\big(96m\max\{B_j, B_{j+1}\}\big)$, $\eps_{j,2} = \Delta/\big(96m \max\{B_j, B_{j+1}\}\big)$\;
For $j\in [m]$, let $W_j$ be the QSVT unitary $U_{P_j}^{(\mathrm{SV})}$ in \thm{qsvt} for the polynomial $P_j$ and the projected unitary encoding $(\Pi_H, \Pi_H, U_H)$ \label{lin:W-j}\;

\For{$j \leftarrow 1$ \KwTo $m$}{  
    Let $V_j$ be the sequence: apply $D_{j-1}$ on $(\mathsf{a}, \mathsf{b},\mathsf{c},  \mathsf{w})$, then $D_j$ on $(\mathsf{a},  \mathsf{b}, \mathsf{d},\mathsf{w})$, and finally $W_j$ on $(\mathsf{a}, \mathsf{w})$\;\label{lin:discriminators}
    
    Let $\ket{\Psi_{\mathrm{in}}} = \ket{\!+\!+}_{\mathsf{a}, \mathsf{b}} \ket{00}_{\mathsf{c}, \mathsf{d}}\otimes (U_I \ket{\mathbf{0}}_{\mathsf{w}})$ be the initial state\;
    Let  $\Pi_{j} = \begin{cases} 
    \ketbra{+}{+}_{\mathsf{a}} \otimes I_{\mathsf{b},\mathsf{c}} \otimes\ketbra{0}{0}_{\mathsf{d}}\otimes (\Pi_H)_\mathsf{w} & \text{if } j=1 \\
     \ketbra{+}{+}_{\mathsf{a}} \otimes I_{\mathsf{b}}\otimes \ketbra{10}{10}_{\mathsf{c}, \mathsf{d}} \otimes (\Pi_H)_\mathsf{w} & \text{if } j\ge 2
\end{cases}$\;
    
    Estimate the squared norm $v_j = \|\Pi_{j} V_j \ket{\Psi_{\mathrm{in}}} \|^2 $ to additive error $\frac{\Delta}{6m B_j}$ via Amplitude Estimation with upper bound $v_j \leq \overline{v}_j$ and maximum failure probability $\delta/m$ in \cor{two-stage-ae}, denoting the output by $\widetilde{v}_j$\;
    
    $S \leftarrow S + B_j \widetilde{v}_j$\;
}
\Return{$S$}\;
\end{algorithm*}

\subsection{Estimate $P_{\alpha}(\mathbf{p})$ from its known upper bound}

\begin{lemma}
\label{lem:non-integer-large-upperbound}
For fixed $\alpha > 1$, suppose $P_{\alpha}(\mathbf{p}) \leq R \leq 1$ and $R$ is known. For any $\eta \in (0, 1/4)$ and $\delta \in (0, 1/3)$, there exists an algorithm $\mathcal{A}$ that outputs an estimate $\widetilde{P}_{\alpha}(\mathbf{p})$ such that 
\begin{equation}\label{eq:algorithm3-output}
|\widetilde{P}_{\alpha}(\mathbf{p})-P_{\alpha}(\mathbf{p})| \leq \eta R
\end{equation}
with success probability at least $1-\delta$, using $$ \widetilde{\mathcal{O}}\left(\frac{n^{\max\left\{0,\frac1\alpha-\frac12\right\}} }{\eta R^{1/\alpha}}\right)$$ queries to $U_p$ and $U_p^{\dagger}$ in \defn{purified-access}.  
\end{lemma}

\begin{proof}
The pseudocode of our algorithm is presented in \algo{algorithm3}, which primarily follows the same vein as \cite[Algorithm 2]{multilevel}. For convenience of our discussion, we denote $\Delta = \eta R$. For each $i \in [n]$, since $p_i^\alpha \leq P_{\alpha}(\mathbf{p}) \leq R$, we have $p_i \leq R^{1/\alpha}$. Different from the analyses in \cite{multilevel}, here every $\sqrt{p_i}$ lies in the interval $[0, R^{1/2\alpha}]$ ranther than $[0, 1]$, so we set $\varphi_0 = R^{1/2\alpha}$ and $\varphi_j = \varphi_0 / 2^{j}$ for $j = 1, 2, \cdots, m+1$. We choose $m$ such that the contribution of those probabilities below $\varphi_{m}^2$ does not exceed $\Delta/12$:
\begin{equation}\label{eq:truncate}
\sum_{i: p_i \leq \varphi_{m}^2} p_i^\alpha \leq n(\varphi_{m})^{2\alpha} \leq \Delta/12.
\end{equation}  
That is, $\varphi_{m} = \Theta\!\left[(\frac{\Delta}{n})^{(1/2\alpha)}\right]$ and $m = \Theta\!\left[\log(\frac{n}{\Delta})\right]$.

In \lin{discriminators} of \algo{algorithm3}, we isolate in loop $j$ the singular values in $(\frac{\varphi_{j+1}}{2}, \frac{\varphi_{j-1}}{2}]$ via non-destructive singular value discriminators $D_{j-1}$ and $D_j$ defined in \thm{nd-svd}, for each $j \in [m]$. More explicitly, for the $i$-th component of the initial state $|\Psi_{\mathrm{in}}\rangle$ (which is independent of the loop number $j$): 
\begin{equation}\label{eq:component}
|\Psi_{\mathrm{in}}^{(i)}\rangle \triangleq |\!+\!+\rangle_{ab} |00\rangle_{cd} \otimes \sqrt{\frac{p_i}{2}} \big(|\phi_{i,+}\rangle + |\phi_{i,-}\rangle\big)  
\end{equation}
such that $\varphi_m < \sqrt{p_i} \in (\varphi_{j_i}, \varphi_{j_i-1}]$, we have in loop $j_i -1$ (not for those $i$ with $\sqrt{p_i} \in (\varphi_{1}, \varphi_{0}]$):
\begin{equation*}\begin{split} 
\bigg\|& D_{j_i-1}D_{j_i-2}|\Psi_{\mathrm{in}}^{(i)}\rangle - \bigg(\mathrm{i} \sqrt{\frac{p_i}{2}} |\!+\!+\rangle_{ab} |1\rangle_{c} |\xi_{i,+}\rangle_{d} |\phi_{i,+}\rangle \\
& + \mathrm{i} \sqrt{\frac{p_i}{2}} |\!+\!+\rangle_{ab} |1\rangle_{c} |\xi_{i,-}\rangle_{d} |\phi_{i,-}\rangle\bigg) \bigg\|\\
\leq & ~p_i\big(\epsilon_{j_i-2,1} + \epsilon_{j_i-2,2} + \epsilon_{j_i-1,1} + \epsilon_{j_i-1,2}\big)
\end{split}\end{equation*}
and
\begin{equation*}\begin{split}
\bigg\| & W_{j_i-1}D_{j_i-1}D_{j_i-2} |\Psi_{\mathrm{in}}^{(i)}\rangle - \\
& \bigg(\mathrm{i} \sqrt{\frac{p_i}{2}} P_{j_i-1}\!\bigg(\frac{\sqrt{p_i}}{2}\bigg) |\!+\!+\rangle_{ab} |1\rangle_{c} |\xi_{i,+}\rangle_{d} |\phi_{i,+}\rangle \\
& + \mathrm{i} \sqrt{\frac{p_i}{2}} P_{j_i-1}\!\bigg(-\frac{\sqrt{p_i}}{2}\bigg) |\!+\!+\rangle_{ab} |1\rangle_{c} |\xi_{i,-}\rangle_{d} |\phi_{i,-}\rangle\\
& + \big|\!\perp_{j_i-1}^{(i)}\!\big\rangle\bigg) \bigg\| \\
\leq & ~p_i\big(\epsilon_{j_i-2,1} + \epsilon_{j_i-2,2} + \epsilon_{j_i-1,1} + \epsilon_{j_i-1,2}\big), 
\end{split}\end{equation*}
where $\big|\!\perp_{j_i-1}^{(i)}\!\big\rangle$ is some non-normalized component such that $|+\rangle\langle +|_\mathsf{a} \otimes(\Pi_H)_\mathsf{w} \big|\!\perp_{j_i-1}^{(i)}\!\big\rangle = 0$. So we get
\begin{equation}\begin{split}\label{eq:discriminate-1} 
& \bigg| \big\| \Pi_{j_i-1} V_{j_i-1} |\Psi_{\mathrm{in}}^{(i)}\rangle \big\|^2 \\
& - p_i P_{j_i-1}\!\bigg(\frac{\sqrt{p_i}}{2}\bigg)^2 \frac{\big|\langle 0|\xi_{i,+}\rangle\big|^2 + \big|\langle 0|\xi_{i,-}\rangle\big|^2}{2}\bigg| \\
\leq &  ~2p_i\big(\epsilon_{j_i-2,1} + \epsilon_{j_i-2,2} + \epsilon_{j_i-1,1} + \epsilon_{j_i-1,2}\big).
\end{split}\end{equation} 
In loop $j_i$, we have:
\begin{equation*}\begin{split}
\bigg\|& D_{j_i}D_{j_i-1}|\Psi_{\mathrm{in}}^{(i)}\rangle - \bigg(\sqrt{\frac{p_i}{2}}  |\!+\!+\rangle_{ab} |\xi_{i,+}\rangle_{c} |0\rangle_{d} |\phi_{i,+}\rangle \\
& - \sqrt{\frac{p_i}{2}} |\!+\!+\rangle_{ab} |\xi_{i,-}\rangle_{c} |0\rangle_{d} |\phi_{i,-}\rangle\bigg) \bigg\|\\
\leq & ~p_i\big(\epsilon_{j_i-1,1} +\epsilon_{j_i-1,2} + \epsilon_{j_i,1} +\epsilon_{j_i,2}\big)
\end{split}\end{equation*}
and
\begin{equation*}\begin{split}
\bigg\| & W_{j_i}D_{j_i}D_{j_i-1} |\Psi_{\mathrm{in}}^{(i)}\rangle - \\
& \bigg(\sqrt{\frac{p_i}{2}} P_{j_i}\!\bigg(\frac{\sqrt{p_i}}{2}\bigg) |\!+\!+\rangle_{ab} |\xi_{i,+}\rangle_{c} |0\rangle_{d} |\phi_{i,+}\rangle \\
& - \sqrt{\frac{p_i}{2}} P_{j_i}\!\bigg(-\frac{\sqrt{p_i}}{2}\bigg) |\!+\!+\rangle_{ab} |\xi_{i,-}\rangle_{c} |0\rangle_{d} |\phi_{i,-}\rangle\\
& + \big|\!\perp_{j_i}^{(i)}\!\big\rangle\bigg) \bigg\| \leq p_i\big(\epsilon_{j_i-1,1} +\epsilon_{j_i-1,2} + \epsilon_{j_i,1} +\epsilon_{j_i,2}\big), 
\end{split}\end{equation*}
where $\big|\!\perp_{j_i}^{(i)}\!\big\rangle$ is some non-normalized component such that $|+\rangle\langle +|_\mathsf{a} \otimes(\Pi_H)_\mathsf{w} \big|\!\perp_{j_i}^{(i)}\!\big\rangle = 0$. So we get
\begin{equation}\begin{split}\label{eq:discriminate-2} 
& \bigg| \big\| \Pi_{j_i} V_{j_i} |\Psi_{\mathrm{in}}^{(i)}\rangle \big\|^2 \\
& - p_i P_{j_i}\!\bigg(\frac{\sqrt{p_i}}{2}\bigg)^2 \frac{\big|\langle 1|\xi_{i,+}\rangle\big|^2 + \big|\langle 1|\xi_{i,-}\rangle\big|^2}{2}\bigg| \\
\leq & ~2p_i\big(\epsilon_{j_i-1,1} +\epsilon_{j_i-1,2} + \epsilon_{j_i,1} +\epsilon_{j_i,2}\big).
\end{split}\end{equation}

For $j \notin \{j_i-1, j_i\}$, after acting $V_j$ on $|\Psi_{\mathrm{in}}^{(i)}\rangle$ in loop $j$, either the register $c$ becomes $|0\rangle$ or the register $d$ becomes $|1\rangle$ approximately, therefore 
\begin{equation}\begin{split}\label{eq:discriminate-3}  
\left\|\Pi_j V_j |\Psi_{\mathrm{in}}^{(i)}\rangle\right\|^2 \leq 2p_i \big(\epsilon_{j-1,1} +\epsilon_{j-1,2} + \epsilon_{j,1} +\epsilon_{j,2}\big).
\end{split}\end{equation}

We choose each local polynomial $P_j$ such that $B_j P_j(\frac{\sqrt{p}}{2})^2$ approximates $p^{\alpha-1}$ up to error $\Delta/12$ for any $\sqrt{p} \in (\varphi_{j+1}, \varphi_{j-1}]$. To see this, let $x = \sqrt{p}/2$, then we have
\begin{equation}\begin{split}\label{eq:polynomial-P}
 & \bigg|B_{j}P_{j}\big(x\big)^2 - (2x)^{2(\alpha - 1)}\bigg| \\
= & B_{j} \left|P_{j}\big(x\big)^2 - \frac{(2x)^{2(\alpha - 1)}}{2^{2\alpha}\varphi_{j-1}^{2(\alpha-1)}}\right| \\
\leq & B_{j} \left|P_{j}\big(x\big) - \frac{(2x)^{\alpha - 1}}{2^\alpha\varphi_{j-1}^{\alpha-1}}\right| \cdot \left|P_{j}\big(x\big) + \frac{(2x)^{\alpha - 1}}{2^\alpha\varphi_{j-1}^{\alpha-1}}\right| \\
\leq & 3B_{j} \left|P_{j}\big(x\big) - \frac{x^{\alpha - 1}}{2^\alpha (\varphi_{j-1}/2)^{\alpha-1}}\right| \leq \Delta/12,
\end{split}\end{equation}
with the last two inequalities coming from properties of $P_j= S_{\alpha-1, \varphi_{j+1}/2, \varphi_{j-1}/2, \Delta/(36B_j)}$ in \lem{positive-powers}. With these settings, we can now show that the output $\sum_{j\in[m]}B_j \widetilde{v}_j$ of \algo{algorithm3} approximates $\sum_{i=1}^{n} p_i^\alpha$ well, up to additional errors of:
\begin{itemize}
\item the cut-off error from \eq{truncate};
\item the amplitude estimation errors in each loop $j$;
\item the leakage errors of \eq{discriminate-1} -- \eq{discriminate-3} from the non-destructive discriminators.
\end{itemize}
To do this, we first prove that: 
\begin{lemma}\label{lem:error-leakage-polynomial}
For the initial state $|\Psi_{\mathrm{in}}\rangle$ and each $V_j$, $\Pi_{j}$, $B_j$ for $j \in [m]$ defined in \algo{algorithm3}, we have 
\begin{equation}\begin{split}\label{eq:error-leakage-polynomial}
& \bigg|\sum_{j \in [m]} B_j \big\|\Pi_{j} V_{j} |\Psi_{\mathrm{in}}\rangle\big\|^2 - \sum_{i: \sqrt{p_i} > \varphi_m} p_i^\alpha\bigg| < 3\Delta/4.
\end{split}\end{equation}
\end{lemma}

The proof of \lem{error-leakage-polynomial} is deferred to the Appendix. In fact, \lem{error-leakage-polynomial} sums up the leakage errors of the non-destructive discriminators and the errors coming from polynomial approximations, which might be the only nontrivial part of the error analysis. For the other errors, it's routine to check that:
\begin{equation}\begin{split}
& \left|\widetilde{P}_\alpha(\mathbf{p})-\sum_{i=1}^n p_i^\alpha\right| \\
\leq & \sum_{i: \sqrt{p_i} \leq \varphi_m} p_i^\alpha + \bigg|\widetilde{P}_\alpha(\mathbf{p})-\sum_{i: \sqrt{p_i} > \varphi_m} p_i^\alpha\bigg| \\
= & \sum_{i: \sqrt{p_i} \leq \varphi_m} p_i^\alpha + \bigg|\sum_{j \in [m]} B_j \widetilde{v}_j - \sum_{i: \sqrt{p_i} > \varphi_m} p_i^\alpha\bigg|  \\
\leq & \sum_{i: \sqrt{p_i} \leq \varphi_m} p_i^\alpha + \sum_{j \in [m]} B_j \bigg|\widetilde{v}_j - \big\|\Pi_{j} V_{j} |\Psi_{\mathrm{in}}\rangle\big\|^2 \bigg| \\
& + \bigg|\sum_{j \in [m]} B_j \big\|\Pi_{j} V_{j} |\Psi_{\mathrm{in}}\rangle\big\|^2 - \sum_{i: \sqrt{p_i} > \varphi_m} p_i^\alpha\bigg| \\
< & ~\Delta/12 + \sum_{j \in [m]} B_j \frac{\Delta}{6m B_j} +  3\Delta/4 = \Delta,\\
\end{split}\end{equation}
so the output of \algo{algorithm3} satisfies \eq{algorithm3-output}. The total success probability of \algo{algorithm3} is at least $1-\delta$, due to the union bound. 

Next we deal with the complexity of \algo{algorithm3}. Fix the loop number $j \in [m]$ and denote $N_j = \left|\big\{i: \sqrt{p_i} \in (\varphi_{j+1}, \varphi_{j-1}]\big\}\right|$, then since 
\begin{equation}
N_j \varphi_{j+1}^{2\alpha} < \sum_{i: \sqrt{p_i} \in (\varphi_{j+1}, \varphi_{j-1}]} p_i^\alpha \leq R,
\end{equation}
we have $N_j \leq \min\big\{n, R/\varphi_{j+1}^{2\alpha}\big\} \triangleq \overline{N}_j$. By \eq{discriminate-1} -- \eq{discriminate-3}, we have 
\begin{equation}\begin{split}
v_j = & \big\| \Pi_{j} V_{j}  |\Psi_{\mathrm{in}}\rangle\big\|^2 \\
\leq & \sum_{\sqrt{p_i} \in (\varphi_{j+1}, \varphi_{j-1}]} p_i + 2\big(\epsilon_{j-1,1} + \epsilon_{j-1,2} + \epsilon_{j,1} + \epsilon_{j,2} \big)\\
\leq & \min\left\{1, \frac{5\overline{N}_j \varphi_{j-1}^2}{4}\right\} = \overline{v}_j.
\end{split}\end{equation}
Therefore, up to logarithmic factors, the amplitude amplification in loop $j$ has query complexity 
\begin{equation}
Q_j = \widetilde{\mathcal{O}}\!\left[\frac{1}{\varphi_j}\left(\frac{B_j \varphi_{j-1}\sqrt{\overline{N}_j}}{\Delta} + \sqrt{\frac{B_j}{\Delta}}\right)\right],
\end{equation}
with the factor $\widetilde{\mathcal{O}}\big(\frac{1}{\varphi_j}\big)$ coming from implemention of the non-destructive singular value discriminators and the QSVT unitary. The total complexity of \algo{algorithm3} is then
\begin{equation}\begin{split}\label{eq:alg3-complexity}
& \widetilde{\mathcal{O}}\!\left[\sum_{j=1}^m \frac{1}{\varphi_j}\left(\frac{B_j \varphi_{j-1}\sqrt{\overline{N}_j}}{\Delta} + \sqrt{\frac{B_j}{\Delta}}\right)\right], \\
= & \widetilde{\mathcal{O}}\!\left[\sum_{j=1}^m \left(\frac{ \min\big\{\varphi_j^{2(\alpha-1)}\sqrt{n}, \varphi_j^{\alpha-2}\sqrt{R}\big\}}{\Delta} + \frac{\varphi_j^{\alpha-2}}{\sqrt{\Delta}}\right)\right]
\end{split}
\end{equation}
by substituting $\overline{N}_j = \min\big\{n, R/\varphi_{j+1}^{2\alpha}\big\}$ and $B_j = 2^{2\alpha}\varphi_{j-1}^{2\alpha-2}$.

For $1 < \alpha <2$, $\varphi_j^{2(\alpha-1)}\sqrt{n}$ increase with $\varphi_j$ while $\varphi_j^{\alpha-2}\sqrt{R}$ decrease with $\varphi_j$, crossing at value $\varphi_j = \big(\frac{R}{n}\big)^{1/(2\alpha)}$. Therefore, $\min\big\{\varphi_j^{2(\alpha-1)}\sqrt{n}, \varphi_j^{\alpha-2}\sqrt{R}\big\} \leq R^{1-\frac{1}{\alpha}}/n^{\frac12-\frac{1}{\alpha}}$ and the first term of \eq{alg3-complexity} sums up to $\widetilde{\mathcal{O}}\big(\frac{R^{1-\frac{1}{\alpha}} n^{\frac{1}{\alpha}-\frac12}}{\Delta}\big)$. The second term is a geometric progression so it sums up to $\widetilde{\mathcal{O}}\big(\frac{\varphi_m^{\alpha-2}}{\sqrt{\Delta}}\big) = \widetilde{\mathcal{O}}\big(\frac{n^{\frac{1}{\alpha}-\frac12}}{\Delta^{\frac{1}{\alpha}}}\big) \leq \widetilde{\mathcal{O}}\big(\frac{R^{1-\frac{1}{\alpha}}n^{\frac{1}{\alpha}-\frac12}}{\Delta}\big)$, where we use $\Delta \leq R/8$.

For $\alpha \geq 2$, the first term is upper-bounded by $\widetilde{\mathcal{O}}\big(\frac{\varphi_0^{\alpha-2}}{\Delta}\big) = \widetilde{\mathcal{O}}\big(\frac{R^{1-\frac{1}{\alpha}}}{\Delta}\big)$. The second term is also a geometric progression and it sums up to $\widetilde{\mathcal{O}}\big(\frac{\varphi_0^{\alpha-2}}{\sqrt{\Delta}}\big) = \widetilde{\mathcal{O}}\big(\frac{R^{\frac12-\frac{1}{\alpha}}}{\sqrt{\Delta}}\big) \leq \widetilde{\mathcal{O}}\big(\frac{R^{1-\frac{1}{\alpha}}}{\Delta}\big)$.

Substituting $\Delta = \eta R$, the total query complexity is therefore 
\begin{equation*}
\widetilde{\mathcal{O}}\left(\frac{n^{\max\{0, \frac{1}{\alpha}-\frac12\}}}{\eta R^{1/\alpha}}\right)
\end{equation*}
as stated.
\end{proof}

\subsection{Estimate $P_{\alpha}(\mathbf{p})$ via binary search}
Using \algo{algorithm3} it self, we can remove the requirement of a known upper bound of $P_\alpha(\mathbf{p})$ through binary search. Since
\begin{equation}\label{eq:large-power-sum-range}
n^{1-\alpha}\leq P_\alpha(\mathbf{p})\leq1,
\end{equation}
we start with $R=1$ and recurently halve $R$ until it reach constant relative accuracy of $P_\alpha(\mathbf{p})$.

\begin{algorithm*}[ht]
\caption{Quantum algorithm for estimating $H_\alpha(\mathbf{p})$ for non-integer $\alpha>1$}
\label{algo:non-integer-large-search}
\Input{Additive error $\epsilon\in(0,1)$; Maximum failure probability $\delta\in(0,1)$; Purified access oracle $U_p$ for $\mathbf{p}=(p_i)_{i=1}^n$\;}
\Parameters{$L=\lfloor(\alpha-1)\log_2n\rfloor$ and $\delta_0=\delta/(L+2)$\;}
\Output{Estimate $\widetilde{H}_\alpha(\mathbf{p})$ of $H_\alpha(\mathbf{p})$ to additive error $\epsilon$.}
\BlankLine
$R\leftarrow1$; $k\leftarrow0$; $\mathsf{stop}\leftarrow0$\;
\While{$\mathsf{stop}=0$}{
Use \algo{algorithm3} with upper bound $R$, multiplicative accuracy $1/8$, and failure probability $\delta_0$ to obtain $\widetilde{P}_\alpha(\mathbf{p})$\;
\eIf{$\widetilde{P}_\alpha(\mathbf{p}) > 3R/8$}{
$\mathsf{stop}\leftarrow1$\;
}{
\If{$k=L$}{\Return{$\mathsf{Failed}$}\;}
$R\leftarrow R/2$; $k\leftarrow k+1$\;
}
}
Use \algo{algorithm3} with the resulting $R$, multiplicative accuracy $\eta=\min\{1/8,(\alpha-1)\epsilon/8\}$, and failure probability $\delta_0$ to obtain $\widetilde{P}_\alpha(\mathbf{p})$\;
\Return{$\widetilde{H}_\alpha(\mathbf{p})=\frac{1}{1-\alpha}\log\widetilde{P}_\alpha(\mathbf{p})$}\;
\end{algorithm*}

\begin{proof}[Proof of \thm{non-integer-large}]
The pseudocode of our algorithm is presented in \algo{non-integer-large-search}. There are $L+1 = \lceil(\alpha-1)\log_2n\rceil$ search steps, with the algorithm returning $\mathsf{Failed}$ if it doesn't stop at the $(L+1)$-th step. A final high-accuracy step is used to present estimate $\widetilde{H}_\alpha(\mathbf{p})$ within the required addictive error $\epsilon$. Since there are at most $L+2$ calls to \algo{algorithm3}, the total falure probability is at most $(L+2)\delta_0 = \delta$. 

In each of the search steps, whenever $\mathsf{stop}=0$, we have $P_\alpha(\mathbf{p})<3R/8+R/8=R/2$. So the update $R\leftarrow R/2$ preserves the input requirement of \algo{algorithm3}. When $\mathsf{stop}=1$, we have $P_\alpha(\mathbf{p})>3R/8-R/8=R/4$ and so
\begin{equation}\label{eq:large-coarse-estimate}
P_\alpha(\mathbf{p})\leq R<4P_\alpha(\mathbf{p}).
\end{equation}

For the final call to \algo{algorithm3} at the high-accuracy step, we set $\eta=\min\{1/8,(\alpha-1)\epsilon/8\}$. Then \eq{large-coarse-estimate} gives relative error of $\widetilde{P}_\alpha(\mathbf{p})$ as
\begin{equation}\label{eq:large-relative-accuracy}
r:=\frac{\eta R}{P_\alpha(\mathbf{p})} < 4\eta \leq \frac12.
\end{equation}
Since $|\log(1+x)| < 2|x|$ for $|x| < 1/2$, the output estimate of \algo{non-integer-large-search} satisfies
\begin{equation}\begin{split}\label{eq:large-entropy-error}
\big|\widetilde{H}_\alpha(\mathbf{p})-H_\alpha(\mathbf{p})\big|
&=\frac{1}{\alpha-1}
\left|\log\frac{\widetilde{P}_\alpha(\mathbf{p})}{P_\alpha(\mathbf{p})}\right|\\
&<\frac{2r}{\alpha-1} < \frac{8\eta}{\alpha-1} \leq \epsilon.
\end{split}\end{equation}

For constant multiplicative accuracy $1/8$, the query complexity for each of the search steps is $\widetilde{\mathcal{O}}\big(n^{\max\{0,1/\alpha-1/2\}}R^{-1/\alpha}\big)$, which grows geometrically as $R$ is halved. So their sum is bounded by the the final scale, which is 
\begin{equation}
\widetilde{\mathcal{O}}\!\left(
\frac{n^{\max\{0,1/\alpha-1/2\}}}{P_\alpha(\mathbf{p})^{1/\alpha}}
\right)
\end{equation}
by \eq{large-coarse-estimate}. The high-accuracy step has query complexity
\begin{equation}\label{eq:large-instance-complexity}
\widetilde{\mathcal{O}}\!\left(
\frac{n^{\max\{0,1/\alpha-1/2\}}}{\epsilon P_\alpha(\mathbf{p})^{1/\alpha}}
\right)
\end{equation}
and since $P_\alpha(\mathbf{p}) \geq n^{1-\alpha}$, the worst-case query complexity of \algo{non-integer-large-search} is bounded by \eq{non-integer-large-complexity} as stated.
\end{proof}


\section{NON-INTEGER R{\'E}NYI ENTROPY ESTIMATOR ($0<\alpha<1$)}\label{sec:non-integer-small}

\begin{theorem}\label{thm:non-integer-small}
For any fixed $0<\alpha<1$, there exists an algorithm $\mathcal{A}$ such that for any $\delta\in(0,1)$, $\epsilon\in(0,1)$ and probability distribution $\mathbf{p}$ on $[n]$, $\mathcal{A}$ can estimate $H_\alpha(\mathbf{p})$ to within additive error $\epsilon$ with success probability at least $1-\delta$, using
\begin{equation}\label{eq:non-integer-small-complexity}
\widetilde{\mathcal{O}}\!\left(
\frac{n^{1/(2\alpha)}}{\epsilon^{\max\left\{1,\frac{1}{2\alpha}\right\}}}
\right)
\end{equation}
queries to $U_p$ and $U_p^\dagger$ in \defn{purified-access}.
\end{theorem}

For $0<\alpha\leq 1/2$, the query upper bound is $\widetilde{\mathcal{O}}\big((n/\epsilon)^{1/(2\alpha)}\big)$; for $1/2\leq\alpha<1$, it is $\widetilde{\mathcal{O}}\big(n^{1/(2\alpha)}/\epsilon\big)$. Compared with the current best upper bound $\widetilde{\mathcal{O}}\!\left(\min\left\{\frac{n^{1/(2\alpha)}}{\epsilon^{1+1/(2\alpha)}},\frac{n^{1/\alpha-1/2}}{\epsilon^{1/\alpha}}\right\}\right)
$ of \cite[Corollary 1.5]{multilevel}, our algorithm improves the precision dependence without increasing the dimensional exponent.

The main difference from \sec{non-integer-large} is the initial state. Here, we prepare a uniform superposition of labels so that the local polynomials avoid approximating the negative power $p_i^{\alpha-1}$ near zero. More explicitly, the initial state is prepared as 
\begin{equation}\begin{split}\label{eq:prepare-uniform}
U_{\mathrm{unif}} |\mathbf{0}\rangle_\mathsf{w} = & \sum_{i=1}^n\frac{1}{\sqrt{n}}  |1\rangle_E |l_i\rangle_{GBCD} \\
= & \sum_{i=1}^n\frac{1}{\sqrt{2n}} \left(|\phi_{i,+} \rangle - |\phi_{i,-} \rangle \right),
\end{split}\end{equation}
with $|l_i \rangle = |0, 0, i, 0\rangle$ being left singular vectors of the projective unitary enconding defined in \eq{encoding2} and $|\phi_{i,\pm} \rangle$ being the eigenvectors of its Hermitian dilation. The preparation unitary $U_{\mathrm{unif}}$ can be easily implemented without queries to $U_p$ or $U_p^\dagger$.

Similarly as \sec{non-integer-large}, we first construct a weaker algorithm which can estimate $P_{\alpha}(\mathbf{p})$ to within multiplicative accuracy $\eta < 1/4$ given its upper bound, and then remove the requirement of knowing an upper bound.

\subsection{Estimate $P_\alpha(\mathbf{p})$ from its known upper bound}

\begin{algorithm*}[ht]
\caption{Quantum algorithm for estimating $P_{\alpha}(\mathbf{p})$ ($0 < \alpha < 1$) from its known upper bound}
\label{algo:algorithm4}
\Input{Upper bound $R \leq n^{1-\alpha}$; Multiplicative accuracy $\eta \in (0, 1/4)$; Maximum failure probability $\delta \in (0, 1/3)$;
    Purified access oracle $U_p$ for the distribution $\mathbf{p} = (p_i)_{i=1}^n$\;
}
\Registers{Auxiliary qubits $\mathsf{a}, \mathsf{b}, \mathsf{c}, \mathsf{d}$ and workspace $\mathsf{w}$ defined in \eq{prepare-uniform}\;
}
\Parameters{$\Delta := \eta R$; $m := \lceil\frac{1}{2\alpha}\log_2(\frac{12n}{\Delta})\rceil$; Intervals $(\varphi_{m+1}, \varphi_m], (\varphi_m, \varphi_{m-1}], \cdots, (\varphi_1, \varphi_0]$ with $\varphi_0 := 1$ and $\varphi_j := 1 / 2^{j}$ for $j = 1, \cdots, m+1$; $B_{m+1} := 0$ and $B_j := 2^{2\alpha + 2}\varphi_{j-1}^{2\alpha}$, local polynomial $Q_j := S_{c,\nu,\beta,\epsilon}$ with $c := \alpha, \nu := \frac{\varphi_{j+1}}{2}, \beta := \frac{\varphi_{j-1}}{2}, \epsilon := \frac{\Delta}{36nB_j}$ in \lem{positive-powers}, $\overline{v}_j := \min\big\{1, 5R/(4n\varphi_{j+1}^{2\alpha})\big\}$ for each $j \in [m]$\;
}
\Output{
    Estimate $\widetilde{P}_\alpha(\mathbf{p})$ satisfying \eq{algorithm4-output}.
}

\BlankLine
$S \leftarrow 0$\;
Let $(\Pi_H, \Pi_H, U_H)$ be Hermitian dilation of the projected unitary encoding defined in \eq{encoding2} and $U_{\mathrm{unif}}$ be the state preparation unitary defined in \eq{prepare-uniform}\;
Let $D_0 = I$ and for $j\in [m]$, let $D_j$ be the non-destructive singular value discriminator $D$ in \thm{nd-svd} with threshold $\gamma = \varphi_j/2$, gap ratio $\rho = 2$, and precision $\eps_{j,1} = \Delta/\big(96mn \max\{B_j, B_{j+1}\}\}\big)$, $\eps_{j,2} = \Delta/\big(96mn \max\{B_j, B_{j+1}\}\big)$\;
For $j\in [m]$, let $W_j$ be the QSVT unitary $U_{Q_j}^{(\mathrm{SV})}$ in \thm{qsvt} for the polynomial $Q_j$ and the projected unitary encoding $(\Pi_H, \Pi_H, U_H)$ \label{lin:W-j}\;

\For{$j \leftarrow 1$ \KwTo $m$}{
   
    Let $V_j$ be the sequence: apply $D_{j-1}$ on $(\mathsf{a}, \mathsf{b},\mathsf{c},  \mathsf{w})$, then $D_j$ on $(\mathsf{a},  \mathsf{b}, \mathsf{d},\mathsf{w})$, and finally $W_j$ on $(\mathsf{a}, \mathsf{w})$\;
    
    Let $\ket{\Psi_{\mathrm{unif}}} = \ket{\!+\!+}_{\mathsf{a}, \mathsf{b}} \ket{00}_{\mathsf{c}, \mathsf{d}}\otimes (U_{\mathrm{unif}} \ket{\mathbf{0}}_{\mathsf{w}})$ be the initial state\;
    Let  $\Pi_{j} = \begin{cases} 
    \ketbra{+}{+}_{\mathsf{a}} \otimes I_{\mathsf{b},\mathsf{c}} \otimes\ketbra{0}{0}_{\mathsf{d}}\otimes (\Pi_H)_\mathsf{w} & \text{if } j=1 \\
     \ketbra{+}{+}_{\mathsf{a}} \otimes I_{\mathsf{b}}\otimes \ketbra{10}{10}_{\mathsf{c}, \mathsf{d}} \otimes (\Pi_H)_\mathsf{w} & \text{if } j\ge 2
\end{cases}$\;
    
    Estimate the squared norm $v_j = \|\Pi_{j} V_j \ket{\Psi_{\mathrm{in}}} \|^2 $ to additive error $\frac{\Delta}{6mn B_j}$ via Amplitude Estimation with upper bound $v_j \leq \overline{v}_j$ and maximum failure probability $\delta/m$ in \cor{two-stage-ae}, denoting the output by $\widetilde{v}_j$\;
    
    $S \leftarrow S + n B_j \widetilde{v}_j$\;
}
\Return{$S$}\;
\end{algorithm*}

\begin{lemma}\label{lem:non-integer-small-upperbound}
For fixed $0<\alpha<1$, suppose $P_\alpha(\mathbf{p})\leq R\leq n^{1-\alpha}$ and $R$ is known. For any $\eta \in (0, 1/4)$ and $\delta\in(0,1/3)$, there exists an algorithm $\mathcal{A}$ that outputs an estimate $\widetilde{P}_\alpha(\mathbf{p})$ such that
\begin{equation}\label{eq:algorithm4-output}
\big|\widetilde{P}_\alpha(\mathbf{p})-P_\alpha(\mathbf{p})\big| \leq \eta R
\end{equation}
with success probability at least $1-\delta$, using
\begin{equation}\label{eq:non-integer-small-upperbound-cost}
\widetilde{\mathcal{O}}\!\left(\frac{n^{1/(2\alpha)}}{R^{1/(2\alpha)}\eta^{\max\left\{1,\frac{1}{2\alpha}\right\}}}\right)
\end{equation}
queries to $U_p$ and $U_p^\dagger$ in \defn{purified-access}.
\end{lemma}

\begin{proof}
The pseudocode of the algorithm is presented in \algo{algorithm4}. We denote $\Delta = \eta R$ for convenience of discussion. Different from \sec{non-integer-large}, here we set $\varphi_0 = 1$ and $\varphi_j = \varphi_0 / 2^{j}$ for $j = 1, 2, \cdots, m+1$, with $m$ chosen such that the contribution of those probabilities below $\varphi_{m}^2$ does not exceed $\Delta/12$:
\begin{equation}\label{eq:truncate-alphasmall}
\sum_{i: p_i \leq \varphi_{m}^2} p_i^\alpha \leq n(\varphi_{m})^{2\alpha} \leq \Delta/12.
\end{equation}  
So we have $\varphi_{m} = \Theta\!\left[(\frac{\Delta}{n})^{1/(2\alpha)}\right]$ and $m = \Theta\!\left[\log(\frac{n}{\Delta})\right]$.

Just as the case of $\alpha > 1$, in loop $j$ of \algo{algorithm4}, we isolate the singular values in $(\frac{\varphi_{j+1}}{2}, \frac{\varphi_{j-1}}{2}]$ via non-destructive singular value discriminators $D_{j-1}$ and $D_j$ defined in \thm{nd-svd}. To be more explicit, for the $i$-th component of the initial state $|\Psi_{\mathrm{unif}}\rangle$: 
\begin{equation}\label{eq:component-alphasmall}
|\Psi_{\mathrm{unif}}^{(i)}\rangle \triangleq |\!+\!+\rangle_{ab} |00\rangle_{cd} \otimes\frac{1}{\sqrt{2n}} \big(|\phi_{i,+}\rangle - |\phi_{i,-}\rangle\big)  
\end{equation}
such that $\varphi_m < \sqrt{p_i} \in (\varphi_{j_i}, \varphi_{j_i-1}]$, we have in loop $j_i -1$ (not for those $i$ with $\sqrt{p_i} \in (\varphi_{1}, \varphi_{0}]$):
\begin{equation*}\begin{split} 
\bigg\|& D_{j_i-1}D_{j_i-2}|\Psi_{\mathrm{unif}}^{(i)}\rangle - \bigg(\mathrm{i}\frac{1}{\sqrt{2n}} |\!+\!+\rangle_{ab} |1\rangle_{c} |\xi_{i,+}\rangle_{d} |\phi_{i,+}\rangle  \\
& - \mathrm{i}\frac{1}{\sqrt{2n}} |\!+\!+\rangle_{ab} |1\rangle_{c} |\xi_{i,-}\rangle_{d} |\phi_{i,-}\rangle\bigg) \bigg\|\\
\leq & ~\frac{1}{n}\big(\epsilon_{j_i-2,1} + \epsilon_{j_i-2,2} + \epsilon_{j_i-1,1} + \epsilon_{j_i-1,2}\big)
\end{split}\end{equation*}
and
\begin{equation*}\begin{split}
\bigg\| & W_{j_i-1}D_{j_i-1}D_{j_i-2} |\Psi_{\mathrm{unif}}^{(i)}\rangle - \\
& \bigg(\mathrm{i}\frac{1}{\sqrt{2n}} Q_{j_i-1}\!\bigg(\frac{\sqrt{p_i}}{2}\bigg) |\!+\!+\rangle_{ab} |1\rangle_{c} |\xi_{i,+}\rangle_{d} |\phi_{i,+}\rangle \\
& - \mathrm{i}\frac{1}{\sqrt{2n}} Q_{j_i-1}\!\bigg(-\frac{\sqrt{p_i}}{2}\bigg) |\!+\!+\rangle_{ab} |1\rangle_{c} |\xi_{i,-}\rangle_{d} |\phi_{i,-}\rangle\\
& + \big|\!\perp_{j_i-1}^{(i)}\!\big\rangle\bigg) \bigg\| \\
\leq & ~\frac{1}{n}\big(\epsilon_{j_i-2,1} + \epsilon_{j_i-2,2} + \epsilon_{j_i-1,1} + \epsilon_{j_i-1,2}\big), 
\end{split}\end{equation*}
where $\big|\!\perp_{j_i-1}^{(i)}\!\big\rangle$ is some non-normalized component such that $|+\rangle\langle +|_\mathsf{a} \otimes(\Pi_H)_\mathsf{w} \big|\!\perp_{j_i-1}^{(i)}\!\big\rangle = 0$. So we get
\begin{equation}\begin{split}\label{eq:discriminate-1-alphasmall} 
& \bigg| \big\| \Pi_{j_i-1} V_{j_i-1} |\Psi_{\mathrm{unif}}^{(i)}\rangle \big\|^2 \\ 
& - \frac{1}{n} Q_{j_i-1}\!\bigg(\frac{\sqrt{p_i}}{2}\bigg)^2 \frac{\big|\langle 0|\xi_{i,+}\rangle\big|^2 + \big|\langle 0|\xi_{i,-}\rangle\big|^2}{2}\bigg| \\
\leq &  ~\frac{2}{n} \big(\epsilon_{j_i-2,1} + \epsilon_{j_i-2,2} + \epsilon_{j_i-1,1} + \epsilon_{j_i-1,2}\big).
\end{split}\end{equation}
In loop $j_i$, we have
\begin{equation*}\begin{split} 
\bigg\|& D_{j_i}D_{j_i-1}|\Psi_{\mathrm{unif}}^{(i)}\rangle - \bigg(\frac{1}{\sqrt{2n}}  |\!+\!+\rangle_{ab} |\xi_{i,+}\rangle_{c} |0\rangle_{d} |\phi_{i,+}\rangle \\
& +\frac{1}{\sqrt{2n}} |\!+\!+\rangle_{ab} |\xi_{i,-}\rangle_{c} |0\rangle_{d} |\phi_{i,-}\rangle\bigg) \bigg\|\\
\leq & ~\frac{1}{n}\big(\epsilon_{j_i-1,1} +\epsilon_{j_i-1,2} + \epsilon_{j_i,1} +\epsilon_{j_i,2}\big)
\end{split}\end{equation*}
and
\begin{equation*}\begin{split}
\bigg\| & W_{j_i}D_{j_i}D_{j_i-1} |\Psi_{\mathrm{unif}}^{(i)}\rangle - \\
& \bigg(\frac{1}{\sqrt{2n}} Q_{j_i}\!\bigg(\frac{\sqrt{p_i}}{2}\bigg) |\!+\!+\rangle_{ab} |\xi_{i,+}\rangle_{c} |0\rangle_{d} |\phi_{i,+}\rangle \\
& +\frac{1}{\sqrt{2n}} Q_{j_i}\!\bigg(-\frac{\sqrt{p_i}}{2}\bigg) |\!+\!+\rangle_{ab} |\xi_{i,-}\rangle_{c} |0\rangle_{d} |\phi_{i,-}\rangle\\
& + \big|\!\perp_{j_i}^{(i)}\!\big\rangle\bigg) \bigg\| \leq \frac{1}{n}\big(\epsilon_{j_i-1,1} +\epsilon_{j_i-1,2} + \epsilon_{j_i,1} +\epsilon_{j_i,2}\big), 
\end{split}\end{equation*}
where $\big|\!\perp_{j_i}^{(i)}\!\big\rangle$ is some non-normalized component such that $|+\rangle\langle +|_\mathsf{a} \otimes(\Pi_H)_\mathsf{w} \big|\!\perp_{j_i}^{(i)}\!\big\rangle = 0$. So we get
\begin{equation}\begin{split}\label{eq:discriminate-2-alphasmall} 
& \bigg| \big\| \Pi_{j_i} V_{j_i} |\Psi_{\mathrm{unif}}^{(i)}\rangle \big\|^2 \\
& - \frac{1}{n} Q_{j_i}\!\bigg(\frac{\sqrt{p_i}}{2}\bigg)^2 \frac{\big|\langle 1|\xi_{i,+}\rangle\big|^2 + \big|\langle 1|\xi_{i,-}\rangle\big|^2}{2} \bigg| \\
\leq & ~\frac{2}{n}\big(\epsilon_{j_i-1,1} +\epsilon_{j_i-1,2} + \epsilon_{j_i,1} +\epsilon_{j_i,2}\big).
\end{split}\end{equation}

For $j \notin \{j_i-1, j_i\}$, after acting $V_j$ on $|\Psi_{\mathrm{unif}}^{(i)}\rangle$ in loop $j$, either the register $c$ becomes $|0\rangle$ or the register $d$ becomes $|1\rangle$ approximately, therefore 
\begin{equation}\begin{split}\label{eq:discriminate-3-alphasmall}  
\left\|\Pi_j V_j |\Psi_{\mathrm{unif}}^{(i)}\rangle\right\|^2 \leq \frac{2}{n} \big(\epsilon_{j-1,1} +\epsilon_{j-1,2} + \epsilon_{j,1} +\epsilon_{j,2}\big).
\end{split}\end{equation}

Here, each local polynomial $Q_j$ is chosen such that $B_j Q_j(\frac{\sqrt{p}}{2})^2$ approximates $p^{\alpha}$ up to error $\Delta/(12n)$ for any $\sqrt{p} \in (\varphi_{j+1}, \varphi_{j-1}]$: letting $x = \sqrt{p}/2$ we have
\begin{equation}\begin{split}\label{eq:polynomial-Q}
 & \bigg|B_{j}Q_{j}\big(x\big)^2 - (2x)^{2\alpha}\bigg| \\
= & B_{j}\left|Q_{j}\big(x\big)^2 - \frac{(2x)^{2\alpha}}{2^{2\alpha+2}\varphi_{j-1}^{2\alpha}}\right| \\
\leq & B_{j}\left|Q_{j}\big(x\big) - \frac{(2x)^{\alpha}}{2^{\alpha+1}\varphi_{j-1}^{\alpha}}\right| \cdot \left|Q_{j}\big(x\big) + \frac{(2x)^{\alpha}}{2^{\alpha+1}\varphi_{j-1}^{\alpha}}\right| \\
\leq & 3B_{j}\left|Q_{j}\big(x\big) - \frac{x^{\alpha}}{2^{\alpha+1} (\varphi_{j-1}/2)^{\alpha}}\right| \leq \frac{\Delta}{12n},
\end{split}\end{equation}
with the last two inequalities coming from properties of $Q_j = S_{\alpha, \varphi_{j+1}/2, \varphi_{j-1}/2, \Delta/(36nB_j)}$ in \lem{positive-powers}. With these settings, we can now show that the output $\sum_{j\in[m]} n B_j \widetilde{v}_j$ of \algo{algorithm4} approximates $\sum_{i=1}^{n} p_i^\alpha$ well, up to additional errors of:
\begin{itemize}
\item the cut-off error from \eq{truncate-alphasmall};
\item the amplitude estimation errors in each loop $j$;
\item the leakage errors of \eq{discriminate-1-alphasmall} -- \eq{discriminate-3-alphasmall} from the non-destructive discriminators.
\end{itemize}
To do this, we first prove that: 
\begin{lemma}\label{lem:error-leakage-polynomial-alphasmall}
For the initial state $|\Psi_{\mathrm{unif}}\rangle$ and each $V_j$, $\Pi_{j}$, $B_j$ for $j \in [m]$ defined in \algo{algorithm4}, we have 
\begin{equation}\begin{split}\label{eq:error-leakage-polynomial-alphasmall}
& \bigg|\sum_{j \in [m]} n B_j \big\|\Pi_{j} V_{j} |\Psi_{\mathrm{unif}}\rangle\big\|^2 - \sum_{i: \sqrt{p_i} > \varphi_m} p_i^\alpha\bigg| < 3\Delta/4.
\end{split}\end{equation}
\end{lemma}
The proof of \lem{error-leakage-polynomial-alphasmall} is deferred to the Appendix. Similarly as \lem{error-leakage-polynomial} in the last section, \lem{error-leakage-polynomial-alphasmall} also sums up the leakage errors of the non-destructive discriminators and the errors coming from polynomial approximations. For the total error, we have:
\begin{equation}\begin{split}
& \left|\widetilde{P}_\alpha(\mathbf{p})-\sum_{i=1}^n p_i^\alpha\right| \\
\leq & \sum_{i: \sqrt{p_i} \leq \varphi_m} p_i^\alpha + \bigg|\widetilde{P}_\alpha(\mathbf{p})-\sum_{i: \sqrt{p_i} > \varphi_m} p_i^\alpha\bigg| \\
= & \sum_{i: \sqrt{p_i} \leq \varphi_m} p_i^\alpha + \bigg|\sum_{j \in [m]} n B_j \widetilde{v}_j - \sum_{i: \sqrt{p_i} > \varphi_m} p_i^\alpha\bigg|  \\
\leq & \sum_{i: \sqrt{p_i} \leq \varphi_m} p_i^\alpha + \sum_{j \in [m]} n B_j \bigg|\widetilde{v}_j - \big\|\Pi_{j} V_{j} |\Psi_{\mathrm{unif}}\rangle\big\|^2 \bigg| \\
& + \bigg|\sum_{j \in [m]} n B_j \big\|\Pi_{j} V_{j} |\Psi_{\mathrm{unif}}\rangle\big\|^2 - \sum_{i: \sqrt{p_i} > \varphi_m} p_i^\alpha\bigg| \\
< & ~\Delta/12 + \sum_{j \in [m]} n B_j \frac{\Delta}{6mn B_j} +  3\Delta/4 = \Delta,\\
\end{split}\end{equation}
so the output of \algo{algorithm4} satisfies \eq{algorithm4-output}. The total success probability of \algo{algorithm4} is at least $1-\delta$, due to the union bound. 

It remains to analyze the query complexity of \algo{algorithm4}. Denote $N_j = \left|\big\{i: \sqrt{p_i} \in (\varphi_{j+1}, \varphi_{j-1}]\big\}\right|$, then since 
\begin{equation}
N_j \varphi_{j+1}^{2\alpha} < \sum_{i: \sqrt{p_i} \in (\varphi_{j+1}, \varphi_{j-1}]} p_i^\alpha \leq R,
\end{equation}
we have $N_j \leq \min\big\{n, R/\varphi_{j+1}^{2\alpha}\big\} \triangleq \overline{N}_j$. By \eq{discriminate-1-alphasmall} -- \eq{discriminate-3-alphasmall}, we have 
\begin{equation}\begin{split}
v_j = & \big\| \Pi_{j} V_{j}  |\Psi_{\mathrm{unif}}\rangle\big\|^2 \\
\leq & \sum_{\sqrt{p_i} \in (\varphi_{j+1}, \varphi_{j-1}]} \frac{1}{n} + 2\big(\epsilon_{j-1,1} + \epsilon_{j-1,2} + \epsilon_{j,1} + \epsilon_{j,2} \big)\\
\leq & \min\left\{1, \frac{5\overline{N}_j }{4n}\right\} = \overline{v}_j.
\end{split}\end{equation}
Therefore, up to logarithmic factors, the amplitude amplification in loop $j$ has query complexity 
\begin{equation}
Q_j = \widetilde{\mathcal{O}}\!\left[\frac{1}{\varphi_j}\left(\frac{B_j \sqrt{n\overline{N}_j}}{\Delta} + \sqrt{\frac{nB_j}{\Delta}}\right)\right],
\end{equation}
with the factor $\widetilde{\mathcal{O}}\big(\frac{1}{\varphi_j}\big)$ coming from implemention of the non-destructive singular value discriminators and the QSVT unitary. The total complexity of \algo{algorithm4} is then
\begin{equation}\begin{split}\label{eq:alg4-complexity}
& \widetilde{\mathcal{O}}\!\left[\sum_{j=1}^m \frac{1}{\varphi_j}\left(\frac{B_j \sqrt{n\overline{N}_j}}{\Delta} + \sqrt{\frac{n B_j}{\Delta}}\right)\right], \\
= & \widetilde{\mathcal{O}}\!\left[\sum_{j=1}^m \left(\frac{ \min\big\{\varphi_j^{2\alpha-1}n, \varphi_j^{\alpha-1}\sqrt{nR}\big\}}{\Delta} + \varphi_j^{\alpha-1}\sqrt{\frac{n}{\Delta}}\right)\right]
\end{split}
\end{equation}
by substituting $\overline{N}_j = \min\big\{n, R/\varphi_{j+1}^{2\alpha}\big\}$ and $B_j = 2^{2\alpha+2}\varphi_{j-1}^{2\alpha}$.

For $1/2<\alpha<1$, $\varphi_j^{2\alpha-1}n$ increases with $\varphi_j$ while $\varphi_j^{\alpha-1}\sqrt{nR}$ decreases with $\varphi_j$, crossing at value $\varphi_j=(\frac{R}{n})^{1/(2\alpha)}$, so $\min\big\{\varphi_j^{2\alpha-1}n, \varphi_j^{\alpha-1}\sqrt{nR}\big\} \leq R^{1-1/(2\alpha)} n^{1/(2\alpha)}$ and the first term of \eq{alg4-complexity} sums up to $\widetilde{\mathcal{O}}\!\left(\frac{R^{1-1/(2\alpha)} n^{1/(2\alpha)}}{\Delta}\right)$. For $0<\alpha \leq 1/2$, the first term of \eq{alg4-complexity} is upper-bounded by $\widetilde{\mathcal{O}}\!\left( \frac{n\varphi_m^{2\alpha-1}}{\Delta} \right) =\widetilde{\mathcal{O}}\!\left[\big(\frac{n}{\Delta}\big)^{1/(2\alpha)}\right]$.

For the second term of \eq{alg4-complexity}, it is a geometric progression for any $0<\alpha<1$, so it sums up to $\widetilde{\mathcal{O}}\left(\sqrt{\frac{n}{\Delta}}\,\varphi_m^{\alpha-1} \right) = \widetilde{\mathcal{O}}\left[\big(\frac{n}{\Delta}\big)^{1/(2\alpha)}\right]$.

Substituting $\Delta = \eta R$, the total query complexity is therefore 
\begin{equation*}
\widetilde{\mathcal{O}}\!\left(\frac{n^{1/(2\alpha)}}{R^{1/(2\alpha)}\eta^{\max\left\{1,\frac{1}{2\alpha}\right\}}}\right)
\end{equation*}
as stated.
\end{proof}

\begin{algorithm*}[ht]
\caption{Quantum algorithm for estimating $H_\alpha(\mathbf{p})$ for $0<\alpha<1$}
\label{algo:non-integer-small-search}
\Input{Additive error $\epsilon\in(0,1)$; Maximum failure probability $\delta\in(0,1)$; Purified access oracle $U_p$ for $\mathbf{p}=(p_i)_{i=1}^n$\;}
\Parameters{$L=\lfloor(1-\alpha)\log_2n\rfloor$ and $\delta_0=\delta/(L+2)$\;}
\Output{Estimate $\widetilde{H}_\alpha(\mathbf{p})$ of $H_\alpha(\mathbf{p})$ to additive error $\epsilon$.}
\BlankLine
$R\leftarrow n^{1-\alpha}$; $k\leftarrow0$; $\mathsf{stop}\leftarrow0$\;
\While{$\mathsf{stop}=0$}{
Use \algo{algorithm4} with upper bound $R$, multiplicative accuracy $1/8$, and failure probability $\delta_0$ to obtain $\widetilde{P}_\alpha(\mathbf{p})$\;
\eIf{$\widetilde{P}_\alpha(\mathbf{p}) > 3R/8$}{
$\mathsf{stop}\leftarrow1$\;
}{
\If{$k=L$}{\Return{$\mathsf{Failed}$}\;}
$R\leftarrow R/2$; $k\leftarrow k+1$\;
}
}
Use \algo{algorithm4} with the resulting $R$, multiplicative accuracy $\eta=\min\{1/8,(1-\alpha)\epsilon/8\}$, and failure probability $\delta_0$ to obtain $\widetilde{P}_\alpha(\mathbf{p})$\;
\Return{$\widetilde{H}_\alpha(\mathbf{p})=\frac{1}{1-\alpha}\log\widetilde{P}_\alpha(\mathbf{p})$}\;
\end{algorithm*}

\subsection{Estimate $P_{\alpha}(\mathbf{p})$ via binary search}
Similarly as the case of $\alpha > 1$, we can remove the requirement of a known upper bound of $P_\alpha(\mathbf{p})$ via binary search, using \algo{algorithm4} it self. Since
\begin{equation}\label{eq:small-power-sum-range}
1\leq P_\alpha(\mathbf{p})\leq n^{1-\alpha},
\end{equation}
we start with $R = n^{1-\alpha}$ and recurently halve $R$ until it reach constant relative accuracy of $P_\alpha(\mathbf{p})$.

\begin{proof}[Proof of \thm{non-integer-small}]
The pseudocode of our algorithm is presented in \algo{non-integer-small-search}. There are $L+1 = \lceil(1-\alpha)\log_2n\rceil$ search steps, with the algorithm returning $\mathsf{Failed}$ if it doesn't stop at the $(L+1)$-th step. A final high-accuracy step is used to present estimate $\widetilde{H}_\alpha(\mathbf{p})$ within the required addictive error $\epsilon$. Since there are at most $L+2$ calls to \algo{algorithm4}, the total falure probability is at most $(L+2)\delta_0 = \delta$. 

In each of the search steps, whenever $\mathsf{stop}=0$, we have $P_\alpha(\mathbf{p})<3R/8+R/8=R/2$. So the update $R\leftarrow R/2$ preserves the input requirement of \algo{algorithm4}. When $\mathsf{stop}=1$, we have $P_\alpha(\mathbf{p})>3R/8-R/8=R/4$ and so
\begin{equation}\label{eq:small-coarse-estimate}
P_\alpha(\mathbf{p})\leq R<4P_\alpha(\mathbf{p}).
\end{equation}

For the final call to \algo{algorithm4} at the high-accuracy step, we set $\eta=\min\{1/8,(\alpha-1)\epsilon/8\}$. Then \eq{small-coarse-estimate} gives relative error of $\widetilde{P}_\alpha(\mathbf{p})$ as
\begin{equation}\label{eq:small-relative-accuracy}
r:=\frac{\eta R}{P_\alpha(\mathbf{p})} < 4\eta \leq \frac12.
\end{equation}
Since $|\log(1+x)| < 2|x|$ for $|x| < 1/2$, the output estimate of \algo{non-integer-small-search} satisfies
\begin{equation}\begin{split}\label{eq:small-entropy-error}
\big|\widetilde{H}_\alpha(\mathbf{p})-H_\alpha(\mathbf{p})\big|
&=\frac{1}{\alpha-1}
\left|\log\frac{\widetilde{P}_\alpha(\mathbf{p})}{P_\alpha(\mathbf{p})}\right|\\
&<\frac{2r}{\alpha-1} < \frac{8\eta}{\alpha-1} \leq \epsilon.
\end{split}\end{equation}

For constant multiplicative accuracy $1/8$, the query complexity for each of the search steps is $\widetilde{\mathcal{O}}\big(n^{1/(2\alpha)}R^{-1/(2\alpha)}\big)$, which grows geometrically as $R$ is halved. So their sum is bounded by the the final scale, which is 
\begin{equation}
\widetilde{\mathcal{O}}\!\left[
\left(\frac{n}{P_\alpha(\mathbf{p})}\right)^{1/(2\alpha)}
\right].
\end{equation}
by \eq{small-coarse-estimate}. The high-accuracy step has query complexity
\begin{equation}\label{eq:small-instance-complexity}
\widetilde{\mathcal{O}}\!\left(\frac{n^{1/(2\alpha)}}{P_\alpha(\mathbf{p})^{1/(2\alpha)}\epsilon^{\max\left\{1,\frac{1}{2\alpha}\right\}}}\right)
\end{equation}
and since $P_\alpha(\mathbf{p}) \geq 1$, the worst-case query complexity of \algo{non-integer-small-search} is bounded by \eq{non-integer-small-complexity} as stated.
\end{proof}

\subsection{lower bound}
For quantum query complexity of estimating R{\'e}nyi entropy under the purified access, \cite[Theorem 2.7]{lifting} gave a lower bound $\Omega\!\left(n^{\frac{1}{2\alpha}-o(1)}+ n^{\frac{1}{2\alpha}-\frac12}/\epsilon^{\frac{1}{2\alpha}}\right)$ for the case of $0<\alpha<1$. Here, we show that a perturbation of a nonzero tail mass gives a stronger lower bound than the construction comparing a point mass with a distribution having a small tail.

\begin{theorem}\label{thm:non-integer-small-lower}
For any fixed $0<\alpha<1$, there exists a constant $\epsilon_\alpha>0$ such that, for all sufficiently large $n$ and any $\epsilon\in(0,\epsilon_\alpha]$, every algorithm that estimates $H_\alpha(\mathbf{p})$ to within additive error $\epsilon$ with success probability at least $2/3$ for every probability distribution $\mathbf{p}$ on $[n]$ requires
\begin{equation}\label{eq:noninteger-small-lower}
\Omega\!\left(
n^{\frac{1}{2\alpha}-o(1)}+
\frac{n^{\frac{1}{2\alpha}-\frac12}}{\epsilon^{\max\left\{1,\frac{1}{2\alpha}\right\}}}
\right)
\end{equation}
queries in the worst case to $U_p$ and $U_p^\dagger$ in \defn{purified-access}, even when their controlled versions are available. 
\end{theorem}

\begin{proof}
It suffices to prove the additional lower bound $\Omega\big(n^{1/(2\alpha)-1/2}/\epsilon\big)$. Consider the family of distributions
\begin{equation}\label{eq:small-tail-family}
\mathbf{p}_\tau=\left(1-\tau,\frac{\tau}{n-1},\ldots,\frac{\tau}{n-1}\right),
\end{equation}
whose power sums satisfy
\begin{equation}\label{eq:small-tail-power}
P_\alpha(\mathbf{p}_\tau)=(1-\tau)^\alpha+(n-1)^{1-\alpha}\tau^\alpha.
\end{equation}
Set $\tau_0=c_\alpha(n-1)^{-(1-\alpha)/\alpha}$, where $c_\alpha\in(0,1/8)$ is a sufficiently small constant. For $\tau\in[\tau_0,2\tau_0]$, we have
\begin{equation*}
c_\alpha^\alpha\leq(n-1)^{1-\alpha}\tau^\alpha\leq(2c_\alpha)^\alpha,
\end{equation*}
whereas $\tau(1-\tau)^{\alpha-1}\leq2c_\alpha(1-2c_\alpha)^{\alpha-1}$. Choose $c_\alpha$ so that the latter is at most $c_\alpha^\alpha/2$. Differentiating $H_\alpha(\mathbf{p}_\tau)$ therefore gives
\begin{equation}\begin{split}\label{eq:small-tail-derivative}
\tau\frac{\d}{\d\tau}H_\alpha(\mathbf{p}_\tau)
&=\frac{\alpha}{1-\alpha}
\frac{(n-1)^{1-\alpha}\tau^\alpha-\tau(1-\tau)^{\alpha-1}}
{P_\alpha(\mathbf{p}_\tau)}\\
&\geq b_\alpha>0,
\end{split}\end{equation}
where $b_\alpha$ is independent of $n$. Let $K_\alpha=6/b_\alpha$ and $\tau_1=(1+K_\alpha\epsilon)\tau_0$. Taking $\epsilon_\alpha$ sufficiently small ensures $K_\alpha\epsilon\leq1$ and hence $\tau_1\leq2\tau_0$. Integrating \eq{small-tail-derivative}, we obtain
\begin{equation}\begin{split}\label{eq:small-tail-separation}
H_\alpha(\mathbf{p}_{\tau_1})-H_\alpha(\mathbf{p}_{\tau_0})
&\geq b_\alpha\log(1+K_\alpha\epsilon)\\
&\geq3\epsilon.
\end{split}\end{equation}
Thus an entropy estimate with additive error $\epsilon$ distinguishes these two distributions with success probability at least $2/3$.

To bound the number of queries, we specify the full unitary oracles. Define the orthonormal states
\begin{equation*}
|h\rangle=|1\rangle_G|1\rangle_B,\qquad
|t\rangle=\frac{1}{\sqrt{n-1}}\sum_{i=2}^n|i\rangle_G|i\rangle_B.
\end{equation*}
Let $V$ be a fixed unitary mapping $|00\rangle_{GB}$ to $|h\rangle$, and let $\mathcal{R}_\theta$ be the real rotation through angle $\theta$ on $\operatorname{span}\{|h\rangle,|t\rangle\}$, acting as the identity on its orthogonal complement. With $\theta_\tau=\arcsin\sqrt{\tau}$, the oracle $U_{\mathbf{p}_\tau}=\mathcal{R}_{\theta_\tau}V$ satisfies
\begin{equation*}
U_{\mathbf{p}_\tau}|00\rangle_{GB}
=\sqrt{1-\tau}|h\rangle+\sqrt{\tau}|t\rangle,
\end{equation*}
and is a valid oracle in \defn{purified-access}. Since $\tau_0\leq\tau_1\leq2\tau_0\leq1/4$, the mean value theorem gives
\begin{equation}\begin{split}\label{eq:small-tail-oracle-distance}
\big\|U_{\mathbf{p}_{\tau_1}}-U_{\mathbf{p}_{\tau_0}}\big\|
&=2\sin\frac{|\theta_{\tau_1}-\theta_{\tau_0}|}{2}\\
&\leq|\theta_{\tau_1}-\theta_{\tau_0}|\\
&=\mathcal{O}(\epsilon\sqrt{\tau_0}).
\end{split}\end{equation}
The same norm bound holds for the inverse and controlled oracles. By the hybrid argument, the trace distance between the final states of any $T$-query algorithm on these two inputs is at most $\mathcal{O}(T\epsilon\sqrt{\tau_0})$. Distinguishing them with success probability at least $2/3$ requires this distance to be bounded below by a positive constant, so
\begin{equation}\label{eq:small-linear-precision-lower}
T=\Omega\!\left(\frac{1}{\epsilon\sqrt{\tau_0}}\right)
=\Omega\!\left(\frac{n^{\frac{1}{2\alpha}-\frac12}}{\epsilon}\right).
\end{equation}
Combining this with \cite[Theorem 2.7]{lifting} proves the claim.
\end{proof}

\thm{non-integer-small-lower}, together with \thm{non-integer-small}, establishes the optimal precision dependence of our R{\'e}nyi entropy estimator for $0 < \alpha < 1$, up to logarithmic factors.


\medskip
\section*{Acknowledgements}
The authors used LLMs for AI-assisted research and language polishing in preparing this paper. In particular, ChatGPT 5.6 contributed to \thm{non-integer-small-lower} for raising the refined lower bound. The authors are responsible for the correctness of all results in this paper.


\medskip

\appendix
\section*{Proof of \lem{error-leakage-polynomial}}
\begin{proof}
We split the l.h.s. of \eq{error-leakage-polynomial} into:
{\allowdisplaybreaks
\begin{align}\label{eq:split}
& \bigg|\sum_{j \in [m]} B_j \big\|\Pi_{j} V_{j} |\Psi_{\mathrm{in}}\rangle\big\|^2 - \sum_{i: \sqrt{p_i} > \varphi_m} p_i^\alpha\bigg| \nonumber\\
\leq & \sum_{i: \sqrt{p_i} > \varphi_m} \bigg|\sum_{j \in [m]} B_j \big\|\Pi_{j} V_{j} |\Psi_{\mathrm{in}}^{(i)}\rangle\big\|^2 - p_i^\alpha \bigg| \nonumber\\
& + \sum_{i: \sqrt{p_i} \leq \varphi_m} \sum_{j \in [m]} B_j \big\|\Pi_{j} V_{j} |\Psi_{\mathrm{in}}^{(i)}\rangle\big\|^2 \nonumber\\
\leq & \sum_{i: \varphi_1 \geq \sqrt{p_i} > \varphi_m} \bigg|\sum_{j \in \{j_i-1, j_i\}} B_j \big\|\Pi_{j} V_{j} |\Psi_{\mathrm{in}}^{(i)}\rangle\big\|^2 - p_i^\alpha \bigg| \nonumber\\
& + \sum_{i: \varphi_1 \geq \sqrt{p_i} > \varphi_m} \sum_{j \notin \{j_i-1, j_i\}} B_j \big\|\Pi_{j} V_{j} |\Psi_{\mathrm{in}}^{(i)}\rangle\big\|^2 \nonumber\\
& + \sum_{i: \sqrt{p_i} \in (\varphi_1, \varphi_0]} \bigg| B_1 \big\|\Pi_{1} V_{1} |\Psi_{\mathrm{in}}^{(i)}\rangle\big\|^2 - p_i^\alpha \bigg| \nonumber\\
& + \sum_{i: \sqrt{p_i} \in (\varphi_1, \varphi_0]} \sum_{2 \leq j \leq m} B_j \big\|\Pi_{j} V_{j} |\Psi_{\mathrm{in}}^{(i)}\rangle\big\|^2 \nonumber\\
& + \sum_{i: \sqrt{p_i} \in (\varphi_{m+1}, \varphi_m]} \sum_{j \in [m]} B_j \big\|\Pi_{j} V_{j} |\Psi_{\mathrm{in}}^{(i)}\rangle\big\|^2 \nonumber\\
& + \sum_{i: \sqrt{p_i} \leq \varphi_{m+1}} \sum_{j \in [m]} B_j \big\|\Pi_{j} V_{j} |\Psi_{\mathrm{in}}^{(i)}\rangle\big\|^2. \nonumber\\
\end{align}
}

For the first term of the last inequality of \eq{split}, we get from \eq{discriminate-1}, \eq{discriminate-2} and \eq{polynomial-P} that
{\allowdisplaybreaks
\begin{align}
& \bigg| \sum_{j \in \{j_i-1, j_i\}} B_j \big\|\Pi_{j} V_{j} |\Psi_{\mathrm{in}}^{(i)}\rangle\big\|^2 - p_i^\alpha \bigg| \nonumber\\
= & ~B_{j_i-1} \bigg| \big\| \Pi_{j_i-1} V_{j_i-1} |\Psi_{\mathrm{in}}^{(i)}\rangle \big\|^2 \nonumber\\
& \qquad - p_i P_{j_i-1}\!\bigg(\frac{\sqrt{p_i}}{2}\bigg)^2  \frac{\big|\langle 0|\xi_{i,+}\rangle\big|^2 + \big|\langle 0|\xi_{i,-}\rangle\big|^2}{2} \bigg|\nonumber\\
& + p_i\bigg|B_{j_i-1}P_{j_i-1}\big(\frac{\sqrt{p_i}}{2}\big)^2 - p_i^{\alpha-1} \bigg| \nonumber\\
& \qquad \cdot \frac{\big|\langle 0|\xi_{i,+}\rangle\big|^2 + \big|\langle 0|\xi_{i,-}\rangle\big|^2}{2}  \nonumber\\
& + B_{j_i} \bigg| \big\| \Pi_{j_i} V_{j_i} |\Psi_{\mathrm{in}}^{(i)}\rangle \big\|^2 \nonumber\\
& \qquad -p_i P_{j_i}\!\bigg(\frac{\sqrt{p_i}}{2}\bigg)^2  \frac{\big|\langle 1|\xi_{i,+}\rangle\big|^2 + \big|\langle 1|\xi_{i,-}\rangle\big|^2}{2} \bigg|\nonumber\\
& + p_i\bigg| B_{j_i} P_{j_i}\big(\frac{\sqrt{p_i}}{2}\big)^2 - p_i^{\alpha-1}\bigg| \frac{\big|\langle 1|\xi_{i,+}\rangle\big|^2 + \big|\langle 1|\xi_{i,-}\rangle\big|^2}{2} \nonumber\\
\leq & ~p_i\Delta/12m + p_i\Delta/12m + p_i\Delta/12 \nonumber\\
= & ~\frac{p_i\Delta}{12}(1+\frac{2}{m}), \nonumber\\
\end{align}
}
and so 
\begin{equation}\begin{split}\label{eq:one}
& \sum_{i: \varphi_1 \geq \sqrt{p_i} > \varphi_m} \bigg|\sum_{j \in \{j_i-1, j_i\}} B_j \big\|\Pi_{j} V_{j} |\Psi_{\mathrm{in}}^{(i)}\rangle\big\|^2 - p_i^\alpha \bigg| \\
& \leq \frac{\Delta}{12}(1+\frac{2}{m}).
\end{split}\end{equation}

For the second term of the last inequality of \eq{split}, we get from \eq{discriminate-3} that
\begin{equation}\begin{split}\label{eq:two}
& \sum_{i: \varphi_1 \geq \sqrt{p_i} > \varphi_m} \sum_{j \notin \{j_i-1, j_i\}} B_j \big\|\Pi_{j} V_{j} |\Psi_{\mathrm{in}}^{(i)}\rangle\big\|^2 \\
\leq & \sum_{i: \varphi_1 \geq \sqrt{p_i} > \varphi_m} \sum_{j \notin \{j_i-1, j_i\}} 2B_j p_i \big(\epsilon_{j-1,1} + \cdots + \epsilon_{j,2}\big) \\
\leq & \frac{(m-2)\Delta}{12m}. 
\end{split}\end{equation}

For the third and the fourth terms, the analyses are much the same as the first two, so that we have
\begin{equation}\label{eq:three}
\sum_{i: \sqrt{p_i} \in (\varphi_1, \varphi_0]} \bigg| B_1 \big\|\Pi_{1} V_{1} |\Psi_{\mathrm{in}}^{(i)}\rangle\big\|^2 - p_i^\alpha \bigg| \leq \frac{\Delta}{12}(1 + \frac{1}{m}),
\end{equation}
and 
\begin{equation}\label{eq:four}
\sum_{i: \sqrt{p_i} \in (\varphi_1, \varphi_0]} \sum_{2 \leq j \leq m} B_j \big\|\Pi_{j} V_{j} |\Psi_{\mathrm{in}}^{(i)}\rangle\big\|^2 \leq \frac{(m-1)\Delta}{12m}.
\end{equation}

For the fifth term, we leverage both \eq{truncate} and \eq{discriminate-1}: 
{\allowdisplaybreaks
\begin{align}\label{eq:five}
& \sum_{i: \sqrt{p_i} \in (\varphi_{m+1}, \varphi_m]} \sum_{j \in [m]} B_j \big\|\Pi_{j} V_{j} |\Psi_{\mathrm{in}}^{(i)}\rangle\big\|^2 \nonumber\\
\leq & \sum_{i: \sqrt{p_i} \in (\varphi_{m+1}, \varphi_m]} B_m \big\|\Pi_{m} V_{m} |\Psi_{\mathrm{in}}^{(i)}\rangle\big\|^2 + \frac{(m-1)\Delta}{12m} \nonumber\\
\leq & \sum_{i: \sqrt{p_i} \in (\varphi_{m+1}, \varphi_m]} B_{m} \bigg|\big\|\Pi_{m} V_{m} |\Psi_{\mathrm{in}}^{(i)}\rangle\big\|^2 - p_iP_{m}\big(\frac{\sqrt{p_i}}{2}\big)^2 \nonumber\\
& \cdot \frac{\big|\langle 0|\xi_{i,+}\rangle\big|^2 + \big|\langle 0|\xi_{i,-}\rangle\big|^2}{2}\bigg| + \frac{(m-1)\Delta}{12m} \nonumber\\
& + \sum_{i: \sqrt{p_i} \in (\varphi_{m+1}, \varphi_m]}p_i \bigg|B_{m}P_{m}\big(\frac{\sqrt{p_i}}{2}\big)^2 - p_i^{\alpha -1} \bigg| \nonumber\\
& \cdot \frac{\big|\langle 0|\xi_{i,+}\rangle\big|^2 + \big|\langle 0|\xi_{i,-}\rangle\big|^2}{2} + \sum_{i: \sqrt{p_i} \in (\varphi_{m+1}, \varphi_m]} p_i^{\alpha}\nonumber\\
\leq & ~\frac{\Delta}{12m} + \frac{(m-1)\Delta}{12m} + \frac{\Delta}{12} + \frac{\Delta}{12}\nonumber\\ 
= & ~\frac{\Delta}{4}. \nonumber\\
&
\end{align}
}

For the sixth term, we get from \eq{discriminate-3} that
\begin{equation}\begin{split}\label{eq:six}
& \sum_{i: \sqrt{p_i} \leq \varphi_{m+1}} \sum_{j \in [m]} B_j \big\|\Pi_{j} V_{j} |\Psi_{\mathrm{in}}^{(i)}\rangle\big\|^2 \\
\leq & \sum_{i: \sqrt{p_i} \leq \varphi_{m+1}} \sum_{j \in [m]} 2B_jp_i \big(\epsilon_{j-1,1} + \epsilon_{j-1,2} + \epsilon_{j,1} + \epsilon_{j,2}\big) \\
\leq & \frac{\Delta}{12}.
\end{split}\end{equation}

Summing up \eq{one} -- \eq{six}, we get the inequality \eq{error-leakage-polynomial}.
\end{proof}

\medskip
\section*{Proof of \lem{error-leakage-polynomial-alphasmall}}
\begin{proof}
We split the l.h.s. of \eq{error-leakage-polynomial-alphasmall} into:
{\allowdisplaybreaks
\begin{align}\label{eq:split-alphasmall}
& \bigg|\sum_{j \in [m]} n B_j \big\|\Pi_{j} V_{j} |\Psi_{\mathrm{unif}}\rangle\big\|^2 - \sum_{i: \sqrt{p_i} > \varphi_m} p_i^\alpha\bigg| \nonumber\\
\leq & \sum_{i: \sqrt{p_i} > \varphi_m} \bigg|\sum_{j \in [m]} n B_j \big\|\Pi_{j} V_{j} |\Psi_{\mathrm{unif}}^{(i)}\rangle\big\|^2 - p_i^\alpha \bigg| \nonumber\\
& + \sum_{i: \sqrt{p_i} \leq \varphi_m} \sum_{j \in [m]} n B_j \big\|\Pi_{j} V_{j} |\Psi_{\mathrm{unif}}^{(i)}\rangle\big\|^2 \nonumber\\
\leq & \sum_{i: \varphi_1 \geq \sqrt{p_i} > \varphi_m} \bigg|\sum_{j \in \{j_i-1, j_i\}} nB_j \big\|\Pi_{j} V_{j} |\Psi_{\mathrm{unif}}^{(i)}\rangle\big\|^2 - p_i^\alpha \bigg| \nonumber\\
& + \sum_{i: \varphi_1 \geq \sqrt{p_i} > \varphi_m} \sum_{j \notin \{j_i-1, j_i\}} nB_j \big\|\Pi_{j} V_{j} |\Psi_{\mathrm{unif}}^{(i)}\rangle\big\|^2 \nonumber\\
& + \sum_{i: \sqrt{p_i} \in (\varphi_1, \varphi_0]} \bigg| nB_1 \big\|\Pi_{1} V_{1} |\Psi_{\mathrm{unif}}^{(i)}\rangle\big\|^2 - p_i^\alpha \bigg| \nonumber\\
& + \sum_{i: \sqrt{p_i} \in (\varphi_1, \varphi_0]} \sum_{2 \leq j \leq m} nB_j \big\|\Pi_{j} V_{j} |\Psi_{\mathrm{unif}}^{(i)}\rangle\big\|^2 \nonumber\\
& + \sum_{i: \sqrt{p_i} \in (\varphi_{m+1}, \varphi_m]} \sum_{j \in [m]} nB_j \big\|\Pi_{j} V_{j} |\Psi_{\mathrm{unif}}^{(i)}\rangle\big\|^2 \nonumber\\
& + \sum_{i: \sqrt{p_i} \leq \varphi_{m+1}} \sum_{j \in [m]} nB_j \big\|\Pi_{j} V_{j} |\Psi_{\mathrm{unif}}^{(i)}\rangle\big\|^2. \nonumber\\
\end{align}
}

For the first term of the last inequality of \eq{split-alphasmall}, we get from \eq{discriminate-1-alphasmall}, \eq{discriminate-2-alphasmall} and \eq{polynomial-Q} that
{\allowdisplaybreaks
\begin{align}
& \bigg| \sum_{j \in \{j_i-1, j_i\}} nB_j \big\|\Pi_{j} V_{j} |\Psi_{\mathrm{unif}}^{(i)}\rangle\big\|^2 - p_i^\alpha \bigg| \nonumber\\
= & ~nB_{j_i-1} \bigg| \big\| \Pi_{j_i-1} V_{j_i-1} |\Psi_{\mathrm{in}}^{(i)}\rangle \big\|^2 \nonumber\\
& \qquad - \frac{1}{n} Q_{j_i-1}\!\bigg(\frac{\sqrt{p_i}}{2}\bigg)^2  \frac{\big|\langle 0|\xi_{i,+}\rangle\big|^2 + \big|\langle 0|\xi_{i,-}\rangle\big|^2}{2} \bigg|\nonumber\\
& + \bigg|B_{j_i-1}Q_{j_i-1}\big(\frac{\sqrt{p_i}}{2}\big)^2 - p_i^{\alpha} \bigg| \nonumber\\
& \qquad \cdot \frac{\big|\langle 0|\xi_{i,+}\rangle\big|^2 + \big|\langle 0|\xi_{i,-}\rangle\big|^2}{2}  \nonumber\\
& + nB_{j_i} \bigg| \big\| \Pi_{j_i} V_{j_i} |\Psi_{\mathrm{unif}}^{(i)}\rangle \big\|^2 \nonumber\\
& \qquad -\frac{1}{n} Q_{j_i}\!\bigg(\frac{\sqrt{p_i}}{2}\bigg)^2  \frac{\big|\langle 1|\xi_{i,+}\rangle\big|^2 + \big|\langle 1|\xi_{i,-}\rangle\big|^2}{2} \bigg|\nonumber\\
& + \bigg| B_{j_i} Q_{j_i}\big(\frac{\sqrt{p_i}}{2}\big)^2 - p_i^{\alpha}\bigg| \frac{\big|\langle 1|\xi_{i,+}\rangle\big|^2 + \big|\langle 1|\xi_{i,-}\rangle\big|^2}{2} \nonumber\\
\leq & ~\Delta/12mn + \Delta/12mn + \Delta/12n \nonumber\\
= & ~\frac{\Delta}{12n}(1+\frac{2}{m}),\nonumber\\
\end{align}
}
and so 
\begin{equation}\begin{split}\label{eq:one-alphasmall}
& \sum_{i: \varphi_1 \geq \sqrt{p_i} > \varphi_m} \bigg|\sum_{j \in \{j_i-1, j_i\}} nB_j \big\|\Pi_{j} V_{j} |\Psi_{\mathrm{unif}}^{(i)}\rangle\big\|^2 - p_i^\alpha \bigg| \\
& \leq \frac{\Delta}{12}(1+\frac{2}{m}).
\end{split}\end{equation}

For the second term of the last inequality of \eq{split-alphasmall}, we get from \eq{discriminate-3-alphasmall} that
\begin{equation}\begin{split}\label{eq:two-alphasmall}
& \sum_{i: \varphi_1 \geq \sqrt{p_i} > \varphi_m} \sum_{j \notin \{j_i-1, j_i\}} n B_j \big\|\Pi_{j} V_{j} |\Psi_{\mathrm{unif}}^{(i)}\rangle\big\|^2 \\
\leq & \sum_{i: \varphi_1 \geq \sqrt{p_i} > \varphi_m} \sum_{j \notin \{j_i-1, j_i\}} 2B_j \big(\epsilon_{j-1,1} + \cdots + \epsilon_{j,2}\big) \\
\leq & n(m-2)\frac{\Delta}{12mn} = \frac{(m-2)\Delta}{12m}. 
\end{split}\end{equation}

For the third and the fourth terms, the analyses are much the same as the first two, so that we have
\begin{equation}\label{eq:three-alphasmall}
\sum_{i: \sqrt{p_i} \in (\varphi_1, \varphi_0]} \bigg| nB_1 \big\|\Pi_{1} V_{1} |\Psi_{\mathrm{unif}}^{(i)}\rangle\big\|^2 - p_i^\alpha \bigg| \leq \frac{\Delta}{12}(1 + \frac{1}{m}),
\end{equation}
and 
\begin{equation}\label{eq:four-alphasmall}
\sum_{i: \sqrt{p_i} \in (\varphi_1, \varphi_0]} \sum_{2 \leq j \leq m} nB_j \big\|\Pi_{j} V_{j} |\Psi_{\mathrm{unif}}^{(i)}\rangle\big\|^2 \leq \frac{(m-1)\Delta}{12m}.
\end{equation}

For the fifth term, we leverage both \eq{truncate-alphasmall} and \eq{discriminate-1-alphasmall}: 
{\allowdisplaybreaks
\begin{align}\label{eq:five-alphasmall}
& \sum_{i: \sqrt{p_i} \in (\varphi_{m+1}, \varphi_m]} \sum_{j \in [m]} n B_j \big\|\Pi_{j} V_{j} |\Psi_{\mathrm{unif}}^{(i)}\rangle\big\|^2 \\
\leq & \sum_{i: \sqrt{p_i} \in (\varphi_{m+1}, \varphi_m]} n B_m \big\|\Pi_{m} V_{m} |\Psi_{\mathrm{unif}}^{(i)}\rangle\big\|^2 + \frac{(m-1)\Delta}{12m} \nonumber\\
\leq & \sum_{i: \sqrt{p_i} \in (\varphi_{m+1}, \varphi_m]} n B_{m} \bigg|\big\|\Pi_{m} V_{m} |\Psi_{\mathrm{unif}}^{(i)}\rangle\big\|^2 - \frac{1}{n}Q_{m}\big(\frac{\sqrt{p_i}}{2}\big)^2 \nonumber\\
& \cdot \frac{\big|\langle 0|\xi_{i,+}\rangle\big|^2 + \big|\langle 0|\xi_{i,-}\rangle\big|^2}{2}\bigg| + \frac{(m-1)\Delta}{12m} \nonumber\\
& + \sum_{i: \sqrt{p_i} \in (\varphi_{m+1}, \varphi_m]} \bigg|B_{m}Q_{m}\big(\frac{\sqrt{p_i}}{2}\big)^2 - p_i^{\alpha} \bigg| \nonumber\\
& \cdot \frac{\big|\langle 0|\xi_{i,+}\rangle\big|^2 + \big|\langle 0|\xi_{i,-}\rangle\big|^2}{2} + \sum_{i: \sqrt{p_i} \in (\varphi_{m+1}, \varphi_m]} p_i^{\alpha}\nonumber\\
\leq & ~\frac{\Delta}{12m} + \frac{(m-1)\Delta}{12m} + \frac{\Delta}{12} + \frac{\Delta}{12}\nonumber\\
= & ~\frac{\Delta}{4}. \nonumber\\
\end{align}
}

For the sixth term, we get from \eq{discriminate-3-alphasmall} that
\begin{equation}\begin{split}\label{eq:six-alphasmall}
& \sum_{i: \sqrt{p_i} \leq \varphi_{m+1}} \sum_{j \in [m]} nB_j \big\|\Pi_{j} V_{j} |\Psi_{\mathrm{unif}}^{(i)}\rangle\big\|^2 \\
\leq & \sum_{i: \sqrt{p_i} \leq \varphi_{m+1}} \sum_{j \in [m]} 2B_j \big(\epsilon_{j-1,1} + \epsilon_{j-1,2} + \epsilon_{j,1} + \epsilon_{j,2}\big) \\
\leq & nm\frac{\Delta}{12mn} = \frac{\Delta}{12}.
\end{split}\end{equation}

Summing up \eq{one-alphasmall} -- \eq{six-alphasmall}, we get the inequality \eq{error-leakage-polynomial-alphasmall}.
\end{proof}

\medskip





\end{document}